\documentclass[useAMS,usenatbib]{mn2e}
\usepackage{graphicx,amssym}

\def\gimic{{\sc gimic}}
\newcommand{\bc}{\begin{center}}
\newcommand{\ec}{\end{center}}

\title[Metallicities of simulated galaxies]
      {The evolution of galaxy metallicity scaling relations in cosmological hydrodynamical simulations}
\author[De Rossi et al.]
       {\parbox{17cm}{M.~E. De Rossi$^{1,2}$ \thanks{Email:derossi@iafe.uba.ar}, T. Theuns$^{3}$, A.~S.  Font$^{4}$, I.~G. McCarthy$^{4}$}
       \\     
       \\
       $^{1}$ Consejo Nacional de Investigaciones Cient\'ificas y T\'ecnicas, Argentina\\  
       $^{2}$ Instituto de Astronom\'{\i}a y F\'{\i}sica del Espacio (IAFE, CONICET-UBA)\\ 
       $^{3}$ Institute for Computational Cosmology, Department of Physics, University of Durham\\
       $^{4}$ Astrophysics Research Institute, Liverpool John Moores University
}
\begin{document}

\date{Accepted  ???? ??. 2010 ???? ??}

\pagerange{\pageref{firstpage}--\pageref{lastpage}} 
\pubyear{2013}

\maketitle

\label{firstpage}

\begin{abstract}
The evolution of the metal content of galaxies and its relations to other global properties [such as total stellar
mass ($M_*$), circular velocity, star formation rate (SFR), halo mass, etc.] provides important constraints
on models of galaxy formation. Here we examine the evolution of metallicity scaling relations of simulated
galaxies in the Galaxies-Intergalactic Medium Interaction Calculation suite of cosmological simulations.
We make comparisons to observations of the correlation of gas-phase abundances with $M_*$ (the
mass-metallicity relation, MZR), as well as with both $M_*$ and SFR or gas mass fraction (the so-called
3D fundamental metallicity relations, FMRs). The simulated galaxies follow the observed local MZR and
FMRs over an order of magnitude in $M_*$, but overpredict the metallicity of massive galaxies ($\log M_*
\ga 10.5$), plausibly due to inefficient feedback in this regime. We discuss the origin of the MZR and
FMRs in the context of galactic outflows and gas accretion. We examine the evolution of mass-metallicity
relations defined using different elements that probe the three enrichment channels (SNII, SNIa, and AGB
stars). Relations based on elements produced mainly by SNII evolve weakly, whereas those based on
elements produced preferentially in SNIa/AGB exhibit stronger evolution, due to the longer timescales
associated with these channels. Finally, we compare the relations of central and satellite galaxies, finding
systematically higher metallicities for satellites, as observed. We show this is due to the removal of the
metal poor gas reservoir that normally surrounds galaxies and acts to dilute their gas-phase metallicity
(via cooling/accretion onto the disk), but is lost due to ram pressure stripping for satellites.
\end{abstract}


\section{Introduction}
\label{sec:introduction}

The study of elemental abundances in galaxies (\lq metallicities\rq) can provide important information about their star formation histories as well as the different
processes affecting gas evolution such as the balance between gas infall and outflows \citep[e.g.,][]{dave2010,dayal2012,lilly2013}.
Moreover, the determination of correlations between the metal content
of these systems and other properties such as their stellar masses ($M_{*}$) or gas
fractions ($f_{\rm g}$) is crucial for reconstructing the formation history
of galaxy populations and for constraining galaxy formation models.

Galaxies in the local Universe, $z \sim 0$, exhibit a clear correlation between stellar mass and metallicity, such 
that systems with higher $M_{*}$ are more metal-enriched.  
This relationship between $M_{*}$ and metallicity (the `mass-metallicity'
relation, MZR) was first reported by \citet{lequeux1979} for star-forming galaxies, with subsequent observational studies typically using luminosity as a surrogate 
\citep[e.g., ][]{garnett1987, zaritsky1994, lamareille2004}.  Over the last decade or so, improved data quality and larger statistical samples of galaxies have allowed for the determination of the MZR not only in the local Universe \citep[e.g., ][]{tremonti2004, lee2006, kewley2008} but also at higher redshifts up to $z \sim 3.5$
\citep[e.g., ][]{savaglio2005, erb2006, maiolino2008, hayashi2009, lamareille2009, 
mannucci2009, laralopez2010a, moustakas2011,
cresci2012, yabe2012, yuan2013, zahid2014}.  Qualitatively speaking, these studies have typically found declining metallicities (at fixed stellar mass) with increasing redshift, although there are quantitative differences in the reported evolution rates.  

The {\it scatter} in the observed MZR has also been extensively studied recently \citep[e.g.][]{ellison2009, yates2011,
zahid2012, hughes2013} as it encodes information about the dependence of metallicity 
on secondary physical parameters.  For example, \citet{cooper2008} and \citet{ellison2009} have shown that, at fixed stellar mass, galaxies inhabiting denser environments tend to have higher metallicities. 
Furthermore, higher gas-phase metallicities
have been associated with lower
star formation rates \citep[SFRs, e.g., ][]{ellison2008, andrews2013} 
and specific SFRs \citep[e.g., ][]{ellison2008}, 
lower gas fractions \citep[e.g.][]{zhang2009, hughes2013}, 
smaller sizes \citep[e.g., ][]{ellison2008, yabe2013},
higher dust fractions \citep[e.g., ][]{zahid2014} 
and larger $B \-- R$ and $R \-- H$ colours \citep[e.g., ][]{yabe2013}.

The correlation of the scatter in the MZR with other physical quantities suggests that the MZR is a 2D projection of a more fundamental 3D (at least) relation.  For example, \citet{laralopez2010a} and \citet{mannucci2010} have suggested the 3D relation between $M_{*}$, SFR and gas-phase metallicity is more fundamental.  \citet{mannucci2010} in particular analysed the metallicity as a function of $M_{*}$ and SFR defining what they called the `Fundamental Metallicity Relation' (FMR), in analogy with the fundamental plane of elliptical galaxies
introduced by \citet{dressler1987}.  Since this 2D plane in 3D space is not parallel to the
$M_\star - Z$ plane, projecting the FMR onto the $M_\star - Z$ plane introduces scatter.
 
\citet{mannucci2010} defined a new quantity ${\mu}_{\alpha} = \log (M_{*}) - \alpha \log ({\rm SFR})$ 
with the aim
of obtaining the projection of the MZR with the minimum scatter, 
finding the tightest relation at $\alpha \sim 0.32$.  
The FMR does not appear to evolve significantly up until $z \sim 2.5$, after which the metallicity of galaxies of fixed $M_{*}$ and SFR starts to decline with increasing $z$ \citep[e.g., ][]{troncoso2014}. 
Since the FMR consists of a positive correlation
between $M_{*}$ and metallicity at a given SFR and a negative correlation
between SFR and metallicity at a given $M_{*}$, part of the observed evolution of the MZR at $z < 2.5$ 
might be an artefact because 
galaxy populations that inhabit different regions of the FMR are sampled
at different cosmic epochs.  
For example, many high-$z$ surveys are magnitude-limited in the UV, tending to select high-SFR
systems.
As the metal
content of galaxies decreases with SFR at a given $M_{*}$, at least part of the observed evolution
of the MZR may be a consequence of the higher SFRs of galaxies observed at high
$z$.  This is consistent with the recent findings of \citet{stott2013}, who select systems with similar $M_*$ {\it and} SFR to local samples and find no significant evolution out of the MZR out to $z \sim 1.5$.

While the existence of a 3D relation between $M_{*}$, SFR and metallicity
has been verified in recent years by many groups, who have explored 
different redshifts and mass ranges, there is still some debate
about the detailed features of the relation
\citep[e.g., ][]{mannucci2011, cresci2012,  yabe2012,  
cullen2014, henry2013,
laralopez2013a, laralopez2013b, pilyugin2013, stott2013, yabe2013}.  
These discrepancies are generally related to different kinds of systematic uncertainties
and selection biases that affect the observational studies and make the comparison
between them non-trivial.
For example, the use of different metallicity calibrations, methods for stellar mass
determination and the aperture size may significantly influence the shape and zero point of the MZR.  
Moreover, at high $z$, the exploration of a wide parameter
space is more challenging because of the limited sample sizes.
The evolution of the physical conditions of 
star forming regions with cosmic time should also be taken into account as they can 
produce a differential evolution of metallicity indicators and locally
calibrated metallicity relations may not be valid at high $z$ \citep{cullen2014}.  

There is also still some debate about whether the $M_*$--SFR--metallicity relation is truly fundamental.  For example, \citet[][]{bothwell2013} \citep[see also][]{laralopez2013c} proposed that the FMR can be considered a manifestation of a more fundamental
relation between $M_{*}$, HI mass and metallicity (the HI-FMR).
These authors defined
a new parametrization for projecting the MZR in this 3D space:
${\eta}_{\beta} = \log (M_{*}) - \beta \log ( M({\rm HI}) - 9.80)$, obtaining 
the lowest scatter at $\beta = 0.35$.
While the dependence of the metallicity on SFR saturates at high masses,
the anti-correlation with HI mass persists.  Motivated by this work, we explore below 3D trends between $M_*$, metallicity and both SFR and gas mass fractions.  

Various evolutionary scenarios have been proposed to explain the presence and evolution of the MZR \citep[e.g., ][]{tissera2005, brooks2007, derossi2007,
finlator2008, dave2010, peeples2011, dave2012, dayal2012, yates2011, romeo2013}.  
The lower metallicity of low-mass galaxies is generally thought to be a result of more efficient star formation-driven outflows from shallower potential wells
\citep{larson1974, tremonti2004,
dalcanton2007, kobayashi2007}.
However, a {\it downsizing} scenario, with smaller galaxies exhibiting lower star formation efficiencies
and consequently lower chemical enrichment,
can also potentially lead to a correlation between mass and metallicity \citep{brooks2007, mouchine2008, calura2009}.
Moreover, the infall of metal-poor gas from the IGM or through merger events
has been frequently invoked as a key ingredient to reproduce
the observed trends \citep[e.g.][]{koppen1999, dalcanton2004, finlator2008, dave2010}.
In this scenario, the SF process is driven mainly
by metal-poor inflowing gas, and the dilution of metal abundances by adding gas and the increased SFR resulting from a bigger gas reservoir, then naturally leads to an
anti-correlation between SFR and metal mass fraction $Z$.  
It is possible that differences in the morphological
mix of galaxies and the associated changes in star formation efficiencies over time affect the
evolution of the MZR \citep[see e.g.][for a study in this sense]{calura2009}. 

Although inflow and outflow of gas is likely to contribute to establishing the MZR, a complication to theoretical modelling is that the mass-loading of a wind, $\dot M_w/\dot M_\star$, is not necessarily the same as the metal mass loading of a wind, $\dot M_Z/\dot M_\star$. The sub-pc wind simulations presented by \citet{creasey2015} demonstrate that dense galactic disks drive metal enriched winds, whereas lower density disks drive winds with far lower metallicities. Interestingly, \citet{derossi2007} showed that a correlation between metallicity and mass naturally arises in a hierarchical framework as a consequence of the regulation of the SFR by merger events, without the necessity of invoking SN-driven outflows. However, without SN feedback, they were not able to reproduce the observed slope of the relation (and neglect of SN feedback also leads to significant overcooling). 
Finally,  the MZR need not result from the efficiency of winds as function of stellar mass, but might simply reflect variations in the stellar initial mass function (IMF) \citep{koppen2007}. 

Although significant progress has been made both characterising and understanding the nature of the MZR and of the FMR, no clear consensus has emerged for either of them.
In this paper we investigate if current models of galaxy formation and evolution are able to describe 
the observed trends, and if so, why.  To do so we analyse the abundance evolution of star-forming regions 
in galaxies from the Galaxies-Intergalactic Medium Interaction Calculation \citep[\gimic,][]{crain2009}
suite of cosmological hydrodynamical simulations.  
These simulations provide a very
large sample of galaxies allowing extrapolation of
statistics to the (500 Mpc $h^{-1}$)$^3$ Millennium Simulation \citep{springel2005} 
volume. They are able to reproduce
rare systems such as voids and massive clusters and
have been shown to describe well dynamical, kinematical and abundance properties of observed galaxies
\citep[e.g., ][]{font2011, mccarthy2012, sales2012}, and are therefore well suited to the problem at hand.  
We examine galaxy metallicity scaling relations and their evolution in the \gimic\ simulations.  In particular, we make comparisons between \gimic\ and observations of the MZR (and its 3D extensions) of local galaxies, we explore the evolution of metallicity scaling relations with stellar mass and circular velocity, and we examine the difference between central and satellite galaxies in terms of their enrichment histories.
We also analyse the
abundance evolution of different elements.

The plan of the paper is as follows. The simulation and sample of galaxies
are described in Sections \ref{sec:simulation_details} and \ref{sec:sample}, respectively.
The redshift $z=0$ mass-metallicity relation is presented in 
Section \ref{sec:local_mzr} and compared to observations, while in Section \ref{sec:evolution}, we analyse and discuss its evolution. In Section \ref{sec:satellites},
we compare the metal enrichment of central and satellite galaxies in \gimic. 
In Section \ref{sec:discussion}, we examine the dominant processes that
determine the main features of the simulated mass-metallicity relation.
Finally, our conclusions are summarised in Section \ref{sec:conclusions}.

\section{The \gimic\ simulations}
\label{sec:simulation}
\subsection{Cosmological parameters and sub-grid implementation}
\label{sec:simulation_details}

\cite{crain2009} provides full details on the \gimic\ simulations, here we begin by giving a brief overview. Five nearly spherical regions of radius $R\sim 20 \ {\rm Mpc} \ h^{-1}$ where picked at redshift $z=1.5$ from the Millennium simulation \citep{springel2005}, and resimulated using \lq zoomed\rq\ initial conditions and including hydrodynamics. In such zoomed initial conditions, the picked regions are simulated at high resolution, with the rest of the (500 Mpc $h^{-1}$)$^3$ Millennium volume represented much more coarsely with more massive dark matter particles. The five regions were selected to have mean overdensities that deviate by (-2, -1, 0, +1, +2) $\sigma$ from the cosmic mean, where $\sigma$ is the root mean square mass fluctuation on a scale of $\sim \ 20 \ {\rm Mpc} \ h^{-1}$ at $z=1.5$. 
These five regions sample environmentally diverse cosmological regions, ranging from massive clusters to voids. Such a zoom technique allows us to combine the advantages of having high numerical resolution with sampling a range of cosmological diverse environments.

The simulations assumes a cold dark matter universe at present dominated by a cosmological constant, and uses the same values for the cosmological parameters as the Millennium Simulation, $(\Omega_m,\Omega_\Lambda,\Omega_b,\sigma_8,n_s)=(0.25,0.75,0.045,0.9,1)$, where the symbols have their usual meaning. The simulations were performed with the code {\sc gadget-3}, a modified version of {\sc gadget-2} last described by \cite{springel2005a},
and with the sub-grid ingredients for galaxy formation taken from the OverWhelmingly Large Simulations (OWLS) suite \citep{schaye2010} as summarized below. Hydrodynamics in {\sc gadget} code is based on the smoothed particle hydrodynamics (SPH) scheme of \cite{lucy1977} and \cite{gingold1977}, with the cosmic gas represented by a set of particles.

Radiative cooling rates for cosmic gas are computed element-by-element in the presence of
a \citet{haardt2001} ionising UV/X-Ray background as a function of density, temperature and redshift using {\sc CLOUDY} \citep{ferland1998} as described in \citet{wiersma2009a}.
Reionisation of Hydrogen is assumed to take place at redshift $z=9$, and of HeII at redshift $z=3.5$, motivated by observations of the intergalactic medium (e.g., \citealt{theuns2002}).

Star formation is implemented following the prescription of \citet{schaye2008}. Above
a hydrogen number density threshold of $\rho > 0.1 {\rm cm}^{-3}$, gas can self-shield from the UV-background (e.g., \citealt{altay2011}), and becomes thermally unstable \citep{schaye2004}.
Lack of numerical resolution prevents the \gimic\ simulations from following this process in detail, and our sub-grid implementation parametrises the unresolved physics
by increasing the pressure of such star forming gas to $p = p_0\,(\rho/\rho_0)^{4/3}$.
Taking the star formation rate to be the power-law
\begin{equation}
\label{eq:sfrlaw}
\dot\rho_\star=\dot\rho_{\star,0}\,(p/p_0)^n\,,
\end{equation}
yields simulated galaxies that follow the Kennicutt-Schmidt law \citep{kennicutt1998} when in hydrostatic equilibrium, with the constants $\dot\rho_{\star,0}$ and $n$ directly related to the amplitude and slope of the Kennicutt-Schmidt law \citep{schaye2008}. Star formation is then implemented stochastically by converting SPH particles to collisionless \lq star\rq\ particles with a probability set by their star formation rate.
Note that, in the case of the very recent {\sc eagle} simulations, SFR is implemented in a very similar way
but considering a metallicity dependent SF thresdhold
\citep[see][for details]{schaye2015}.

Abundance evolution and enrichment is implemented as described by \citet{wiersma2009b}.
The simulation tracks eleven elements (H, He, C, Ca, N, O, Ne, Mg, S, Si, Fe) 
\footnote{\citet{wiersma2009a} found that these elements contribute significantly 
to the radiative cooling at $T > 10^4$ K.},
produced by three stellar evolutionary channels, ($i)$ core-collapse (type~II) supernovae resulting from massive stars ($M>6M_\odot$), ($ii)$ type I supernovae assumed to result from catastrophic mass transfer in close binary stars, and ($iii$) asymptotic giant branch (AGB) stars. Stellar evolutionary tracks and yields that depend on initial metal abundance
are taken from \citet{portinari1998, marigo2001}; and \citet{thielemann2003}.

Nucleosynthetic yields are uncertain by factors
of a few as discussed in \citet{wiersma2009b} (see also the detailed comparison of 
yields for a range of elements in published models by \citealt{Nomoto13}). 
Apart from uncertainties in the yields of a given element for a given stellar evolutionary channel, there is also discussion in the literature about which channels dominate for a given element. For example, 
\cite{Chiappini06} argue that rotation at low metallicity significantly enhances the production of Nitrogen
and that this is required to solve the primary Nitrogen problem for low-metallicity Milky Way stars whereas
\cite{Kobayashi11}
argue that AGB stars could be the source of the N even at low metallicity as discussed in \cite{Nomoto13}.
A detailed investigation of the abundances of different elements in different environments could help to resolve such issues. For example \cite{pipino2009} shows how the changes in N yields as a function of metallicity that resolve the primary N problem in the Milky Way fails to predict the correct N/Fe ratio in local spheroids. In this paper we use a single set of yield tables and show that abundance patterns of galaxies nevertheless vary significantly between them because the extent to which a galaxy can hang-on to its own metals depends on the efficiency of feedback. This clearly complicates matters and illustrates how one cannot simply compare yield tables to observations and judge which table is the more accurate. Changing yield tables in addition to making changes in the efficiency of feedback  will of course also impact the abundance
evolution.

A star particle in the simulation represents a single stellar population with given age and initial abundance, assuming a  \cite{chabrier2003} IMF. Star particles distribute the synthesised elements and total mass lost during each time-step to neighbouring gas particles according to the SPH interpolation scheme.

Kinetic feedback from SNe is implemented as described by \citet{dallavechia2008}.
After a time delay of $3 \times 10^7 {\rm yr}$, the mean life time of a core collapse progenitor, a star particle \lq kicks\rq\ on average $\eta=4$ gas particles into a wind with launch speed $v_w = 600 {\rm \ km \ s^{-1}}$. The values yield a star formation history that tracks the observations well. The chosen values for the parameters $(\eta,v_w)$ imply that 80 per cent of the available SN energy is used to drive a wind (for a \cite{chabrier2003} IMF with stars forming with masses in the range 0.1-100~$M_\odot$, of which those more massive than $6M_\odot$ undergo core collapse). Energy injected by type~I SNe is also taken into account. It is worth noting that in this scheme, the 'kicked' particles are not hydrodynamically decoupled at any stage.

The \gimic\ simulations have been run at three levels of resolution. The low-resolution version
corresponds to the original Millennium simulation \citep{springel2005}; the intermediate and high resolution correspond to gas particle masses of $1.16 \times 10^7$ and 
$1.45 \times 10^6 h^{-1} M_{\odot}$, respectively, with the dark matter particles a factor of ${\Omega}_{\rm m} - {\Omega}_{\rm b} / {\Omega}_{\rm b} = 4.56$ higher. Gravitational forces are spline-softened with equivalent Plummer-sphere softening length of $\epsilon=0.5h^{-1}$ comoving kiloparsecs for the high resolution run until redshift $z=3$, and is kept constant in physical units thereafter. The intermediate resolution simulation uses $\epsilon=1h^{-1}$~kpc.

For most of the analysis presented below we used the high resolution \gimic\ simulation performed in the -2 and 0$\sigma$ regions. The Appendix compares these results with those from the intermediate resolution runs; we find that the main trends and conclusions are robust against a change in numerical resolution of a factor of eight in mass. Below and unless otherwise stated, we will refer to the mass fraction of elements heavier than Helium as \lq metallicity\rq.

\subsection{Identifying galaxies in \gimic}
\label{sec:sample}

We identify dark matter halos by applying a friends-of-friends (FoF) algorithm with linking length $b=0.2$ times the mean interparticle separation \citep{davis1985}.  We then apply the {\sc subfind} algorithm described in \cite{springel2001} and modified to include baryons by \citet{dolag2009} to identify galaxies as self-bound structures. We will refer to the most massive {\sc subfind} structure of a FoF halo as its \lq central galaxy\rq\ , any other galaxy in the same FoF halo as a \lq satellite\rq. 
Unless otherwise specified, central substructures of FoF-halos are analysed in this work.
To make comparison of simulated galaxies to observations more realistic, we measure properties
of simulated galaxies out to a maximum distance of $R_{\rm cut} = 20 \ {\rm kpc}$ to mimic surface brightness limitations in the data.

We limit our analysis to substructures with more than 2000 particles in total, and following \citet{mccarthy2012}, also require galaxies to have a stellar mass of at least
$M_\star \ge 10^9 M_\odot$ (approximately 500 star particles). We also require
$M_\star\le 10^{10.5} M_\odot$, because \gimic\ does {\em not} include
AGN feedback and as a consequence these more massive galaxies are overcooled and
have too high stellar fractions. These selection cuts result in a final sample of between 450 and 650 simulated galaxies, depending on redshift.
We verified by using the 5 spheres of the intermediate resolution \gimic\ suite that our findings do not depend (strongly) on large-scale-environment and hereafter we do not distinguish between the two different high-resolution \gimic\ spheres (i.e. the -2 and 0$\sigma$ \gimic\ runs). 

Finally, element abundances are stored based on both {\em particle} and SPH {\em smoothed} quantities as explained in \citet{wiersma2009b}. 
The SPH algorithm computes the density $\rho({\bf r}_i)$ at the location ${\bf r_i}$ of particle $i$ by summing the
kernel $W$ over all particles within a smoothing length $h_i$ of ${\bf r}_i$,
\begin{equation}
\rho({\bf r}_i)=\sum_j m_j\,W(|{\bf r}_i­{\bf r}_j|/h_i)\,,
\end{equation}
where $m_j$ is the mass of particle $j$. If $A_j$ is the mass fraction in a given element
for particle $j$ ({\em i.e.} its \lq particle metallicity\rq), then the {\em smoothed metallicity} $\bar A_i$ for element
$A$ of particle $i$ is calculated as
\begin{equation}
\bar A_i\,\rho({\bf r}_i) = \sum_j A_j\,m_j\,W(|{\bf r}_i­{\bf r}_j|/h_i\,).
\end{equation}

Our results do not depend on which of the two abundances we use, for simplicity we only show results corresponding to smoothed abundances below.

\section{Comparison to observed abundance relations}
\begin{figure*}
\begin{center}
\resizebox{17.cm}{!}{\includegraphics{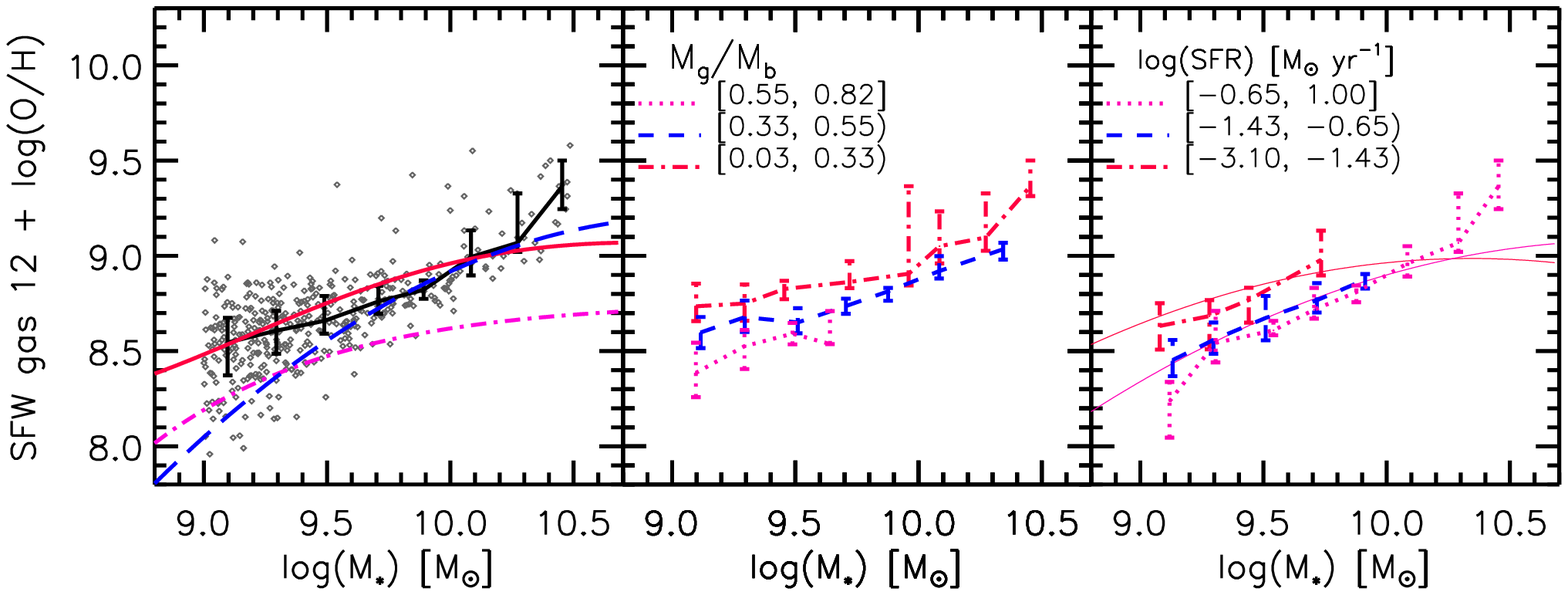}}
\end{center}
\caption[MZR at $z=0$]
{
SFR-weighted gas-phase Oxygen abundance, O/H, as a function of galaxy stellar mass, $M_\star$, at redshift $z=0$. {\em  Left panel:} full \gimic\ sample of central galaxies (small symbols) with mean and 25 and 75th percentiles shown as black solid line with error bars. The {\em  red solid}, {\em blue dashed} and {\em pink dot-dashed} lines are observational fits from \citet{mannucci2010, laralopez2010a} and \citet{sanchez2013}, respectively. There are significant systematic differences between the observational fits, both in amplitude and shape. The simulated abundances are consistent with the data at lower masses but may be slightly higher at the most massive end, $M_\star \approx 10^{10.5} M_\odot$. {\em Middle panel:}
Same as left panel but for \gimic\ galaxies binned in gas fraction, $f_g=M_g/(M_g+M_\star)=M_g/M_b$. At given $M_\star$, gas rich galaxies have lower O/H. {\em Right panel:} same as left panel, but for galaxies binned in star formation rate, $\dot M_\star$. At given $M_\star$, galaxies with higher $\dot M_\star$ have lower O/H, consistent with the observed trend from \citet{mannucci2010}
shown for $\log ({\rm SFR}) = -2$ and 0 as red and pink line, respectively.}
\label{fig:OH_vs_Ms_z0}
\end{figure*}

\begin{figure*}
\begin{center}
\resizebox{9cm}{!}{\includegraphics{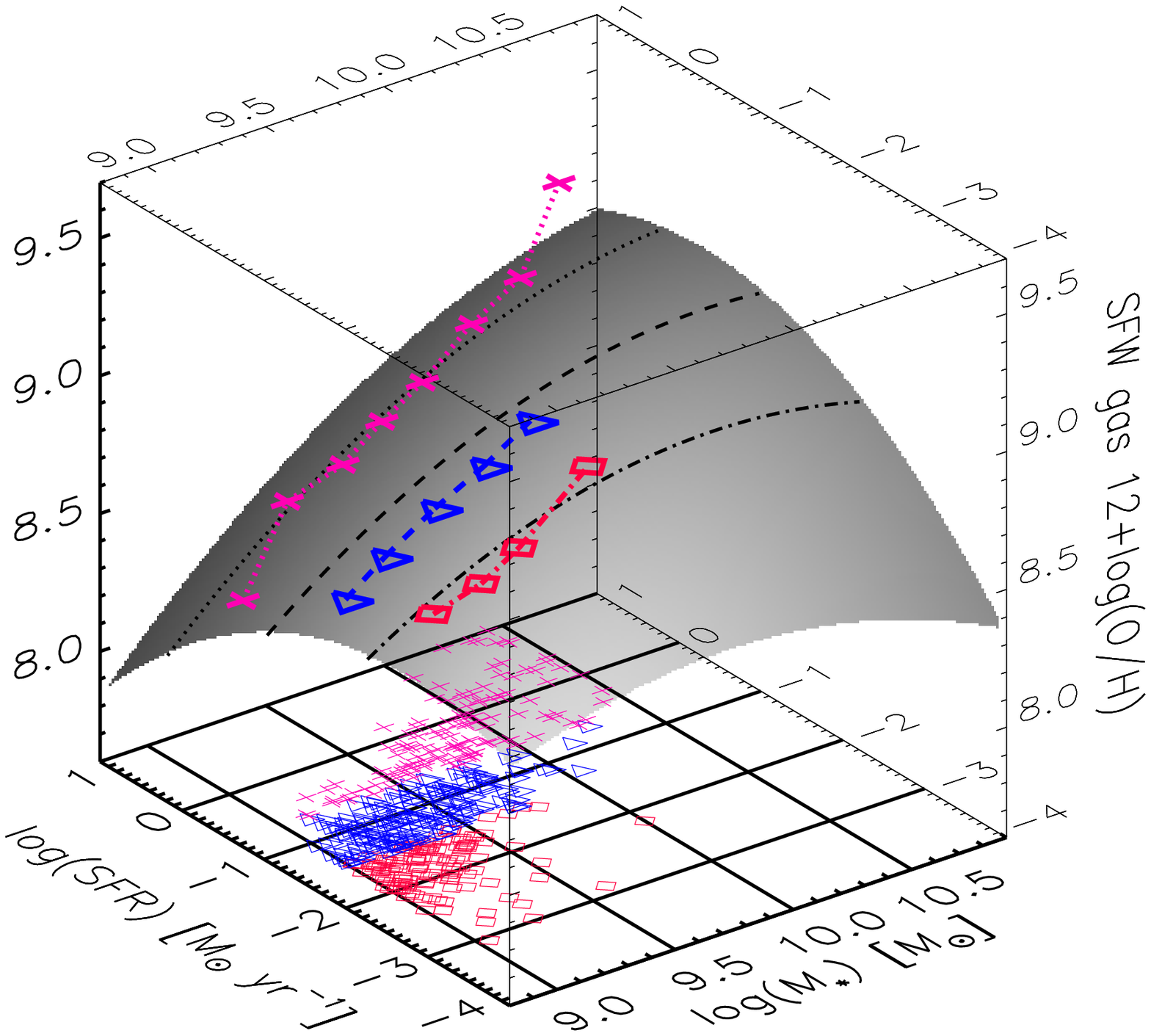}}
\hspace{-1.0cm}\resizebox{9cm}{!}{\includegraphics{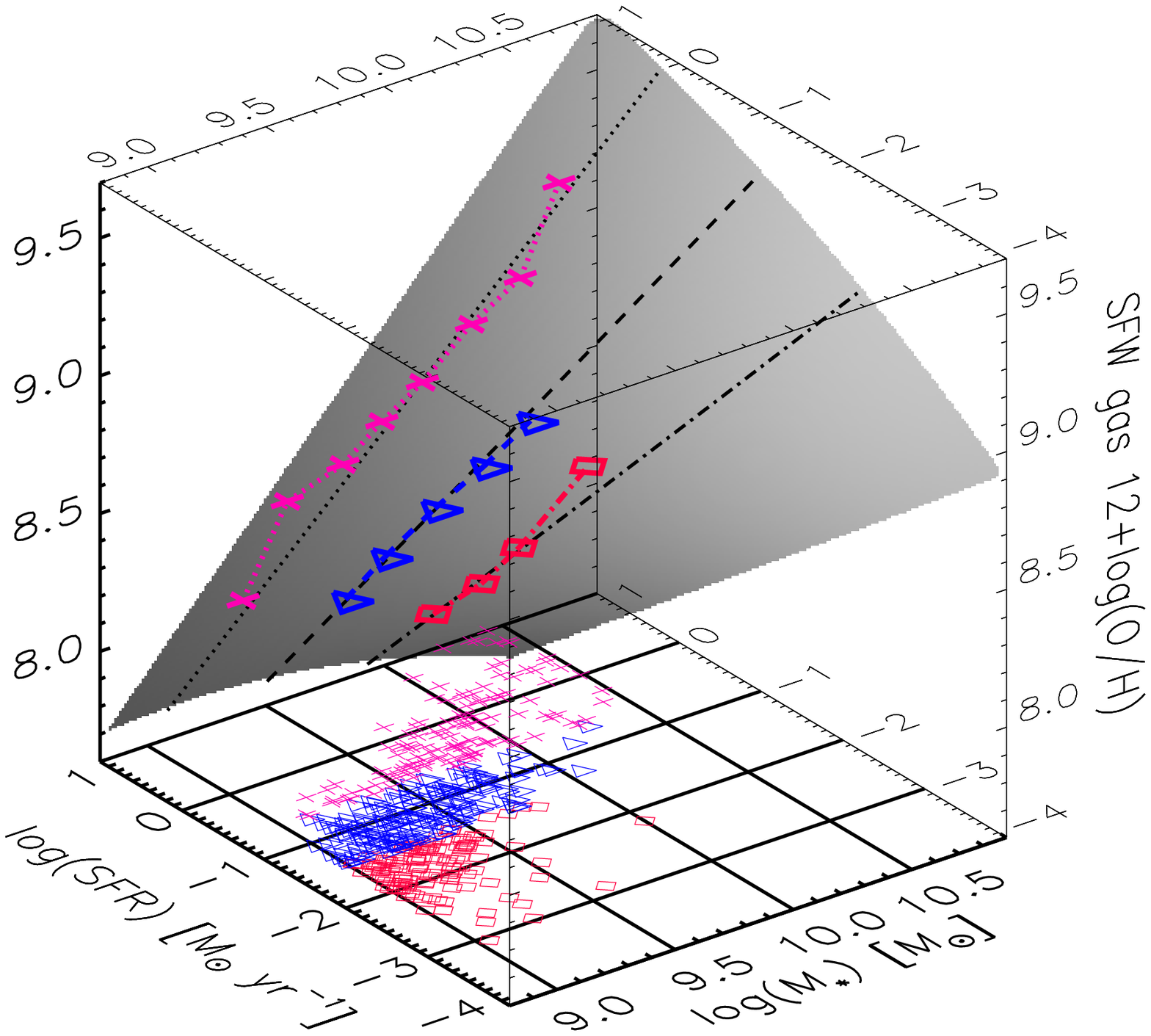}}
\end{center}
\caption[FMR]
{Location of observed galaxies in $M_*$ - SFR - O/H space from \citet{mannucci2010} ({\em left panel}) and of \gimic\ galaxies ({\em right panel}). The shaded area in the left panel represents the best fit 2D surface to a fundamental plane taken from \citet{mannucci2010}, the right panel gives our fit to \gimic\ galaxies. Lines with symbols depict the median $M_\star$-O/H relations for bins in SFR from Fig. \ref{fig:OH_vs_Ms_z0} (right panel), black lines trace the shaded surface at the center of each SFR bin. The relation between $M_\star$ and SFR in the simulations is shown by the symbols in the $M_\star$-SFR plane. 
}
\label{fig:fmr}
\end{figure*}

\subsection{The mass-metallicity relation} 
\label{sec:local_mzr}
The $M_\star$-Z mass metallicity relation (MZR) is a fundamental galaxy scaling relation, linking dynamics in terms of star formation, gas inflow and outflow, to abundances. In this section we compare the MZR in \gimic\ to the observed relation at redshift $z=0$, which is typically inferred by measuring abundances in star forming gas. We will therefore begin by looking at gas-phase abundance in \gimic\, weighing each particle with its star formation rate computed as described in Section~\ref{sec:simulation}. 
The SFR weighted abundance $\bar A$ of an element $A$ of a galaxy is
\begin{equation}
\bar A\,\dot M_\star = \sum_i A_i\dot m_{\star, i}\,,
\end{equation}
where the sum is over all gas particles $i$ in that galaxy, $\dot m_{\star, i}$ and $A_i$ are the star formation rate
and abundance of element $A$ of gas particle $i$, and $\dot M_\star=\sum_i \dot m_{\star, i}$ is the total star
formation rate.
For simulated galaxies we calculate metallicities within the inner ~20kpc, but verified that using 10~kpc instead makes no significant difference.

\gimic\ galaxies exhibit a tight correlation between gas-phase Oxygen abundance weighted by star formation rate, and stellar mass, with more massive systems more metal rich (Fig. \ref{fig:OH_vs_Ms_z0}). The ratio O/H increases by approximately 1~dex over the mass range $10^9-10^{10.5} M_\odot$, with approximately 1~dex of scatter at given $M_\star$.

The observed values of O/H quoted by \citet{mannucci2010} are $\sim 0.4$~dex higher than those of \cite{sanchez2013}, but the dependence on $M_\star$ is very similar, with O/H increasing with $M_\star$ up to $M_\star\sim 10^{10.5} M_\odot$, becoming approximately constant at higher mass. The O/H-$M_\star$ relation reported by \cite{laralopez2010a} is significantly steeper. \gimic\ metallicities follow the O/H-$M_\star$ from \citet{mannucci2010} up to $M_\star\sim 10^{10} M_\odot$, but are higher than those data for more massive galaxies. The absence of AGN feedback in \gimic\ causes overcooling in more massive objects \citep{mccarthy2012}, and the resulting high stellar fractions and lack of feedback may be responsible for the high O/H values.

AGN-feedback plays a crucial role in quenching star formation in massive galaxies 
\citep[e.g.][]{bower2006} and models that do not consider AGN tend to over predict stellar mass and star
formation rates in such galaxies. \citet{schaye2015} present results from the {\sc
EAGLE} \footnote{{\sc eagle} (Evolution and Assembly of GaLaxies and their Environments)
is a project of the Virgo consortium for cosmological supercomputer
simulations.
More information can be found at the {\sc eagle} web
sites at Leiden, http://eagle.strw.leidenuniv.nl/, and Durham,
http://icc.dur.ac.uk/Eagle/
} simulations which reproduce the observed galaxy stellar mass function well and show that
these simulations agree well with the observed MZR also for more massive galaxies (their Fig 13).
We plan to investigate the metal-enrichment history in {\sc EAGLE} in a future work.

\subsection{O/H dependence on SFR and gas fraction}
At a given stellar mass, the O/H abundance of \gimic\ galaxies decreases with increasing gas fraction, $f_g\equiv M_g/(M_\star+M_g)\equiv M_g/M_b$,  and increasing star formation rate (Fig.~\ref{fig:OH_vs_Ms_z0}, middle and right hand panels, respectively). Here, $f_g$ refers to the {\em total} gas fraction - not just the fraction of star forming gas - but we verified that similar trends appear when limiting the gas mass to star forming gas only for lower mass galaxies with $M_\star < 10^{10.3} M_\odot$. The limited sample size restricts us to using a small number of bins in $f_g$ and SFR, and each bin contains 30-40 per cent of the full sample. At the low-mass end of the simulated MZR, the difference in O/H between the highest and lowest $f_{\rm g}$ and SFR bins are $\sim 0.3$~dex and $\sim 0.2$~dex, respectively.

The decrease in O/H with increasing $f_g$ and SFR is also seen in the data of \cite{mannucci2010}
and \cite{bothwell2013}. The solid lines in the right hand panel correspond to the FMR of \citet{mannucci2010}, with red and pink lines corresponding to $\log ({\rm SFR}) = -2$ and 0, respectively. Encouragingly, we find reasonable agreement in zero-point and slope of the simulated relations compared to data up to $M_\star\sim 10^{10} M_\odot$. At higher $M_\star\ge 10^{10.5} M_\odot$ \gimic\ galaxies are too metal rich as noted previously.

In \gimic, SFR is related to the gas reservoir through the imposed (subgrid) star formation law of Eq.~(\ref{eq:sfrlaw}, see Section~\ref{sec:simulation}), which does not simply translate to $\dot M_\star\propto M_g$. However $M_g$ and SFR {\em are} correlated in \gimic\ galaxies, with galaxies with a larger {\em total} gas mass exhibiting higher SFR. By itself, this does not imply a causal connection: the correlation may be a consequence of the same phenomenon, for example an increased gas fraction and enhanced star formation may both result from a recent accretion or merger event that adds gas to the halo. Whatever the origin of the gas, some of it has low elemental abundance
and makes it into the galaxy, since the abundance of star forming gas decreases with increasing SFR. This suggests that at lower stellar mass where these correlations are strongest, accreted low abundance gas causes both the decrease in abundance, and the increase in star formation. The good agreement between the scaling relation in \gimic\ and the observations then suggests that the same process may be at work in real galaxies.

Interestingly, the observed relations turn-over for more massive galaxies, $M\sim 10^{10.5}M_\odot$, in the sense that metallicity no longer increases with $M_\star$. As gas surely still accretes onto the {\em haloes} of these more massive galaxies, the break down of the scaling relation then suggest that the gas no longer accretes onto the galaxy. Contrasting these observed relations to what happens in \gimic\ where the scaling relation does hold for these more massive galaxies suggests the cause of this: feedback. In real galaxies some process prevents the gas from accreting onto the galaxies, but not in \gimic. We suggest it is likely AGN feedback - not included in \gimic\ - that causes the turn-over in the observed MZR.

The location of observed galaxies in O/H, $M_\star$ and SFR space is plotted in Fig.~\ref{fig:fmr}  
({\em left panel}). The correlations discussed so far cause observed galaxies to scatter around a (curved) 2D plane in this 3D space, and the fit of \citet{mannucci2010} to this \lq fundamental plane\rq\,
\begin{eqnarray}
\lefteqn{12 + \log ({\rm O/H}) = 8.90 + 0.37 m - 0.14 s - 0.19 m^2} 
\nonumber\\ & &
{} + 0.12 m s - 0.054 s^2\,,
\end{eqnarray}
is shown in the left panel. Here, $m \equiv \log (M_{*}/10^{10}M_\odot)$ and $s \equiv \log ({\rm SFR}/M_\odot\,{\rm yr}^{-1})$. Similar to the observed galaxies, \gimic\ galaxies also trace a plane in this space, with 
\begin{eqnarray}
\lefteqn{12 + \log ({\rm O/H}) = 8.93 + 0.79 m - 0.047 s + 0.0048 m^2} 
\nonumber\\ & &
{} + 0.19 m s - 0.015 s^2\,.
\end{eqnarray}
The ratio O/H depends more strongly on mass and less strongly on star formation rate in the simulations as compared to the data, but the overall similarity is encouraging.

This \gimic\ {\em fundamental metallicity relation} describes well the median $M_\star$-O/H relation at different SFRs. For both observed and \gimic\ FMRs, O/H is {\em anti}-correlated with SFR at the low-mass end ($M_\star\la 10^9M_\odot$), but this trend reverses for more massive galaxies ($M\star\ga 10^{10}M_\odot$) for which O/H is {\em correlated} with SFR. Such trends are consistent with those found by \cite{yates2011} in semi-analytic model galaxies.

In analogy with the fundamental plane of elliptical galaxies \citep{dressler1987}, it is possible to chose combinations of O/H, SFR and $M_\star$ that minimise the scatter when seeing the FMR edge on, for example using the combination O/H and ${\mu}_{\alpha}\equiv \log(M_\star)-\alpha\,\log({\rm SFR})$. For \gimic, scatter is minimised for a value of $\alpha=0.2$ as compared to $\alpha=0.32$ for the data from \cite{mannucci2010}. Alternatively in terms of $\eta_\beta\equiv \log(M_\star)-\beta\log(M({\rm H\sc I})-9.80)$, \gimic\ galaxies prefer $\beta\approx 0.2$  when using the mass in star forming gas in lieu of $M({\rm H{\sc I}})$, and $\beta=0.48$ when using the total gas mass, as compared to $\beta=0.35$ for observed galaxies according to \cite{bothwell2013}. 
We conclude that the feedback in \gimic\ induces similar correlations between abundance and star formation rate as a function of galaxy stellar mass, although in detail there are differences, especially for more massive galaxies.

\subsection{The $f_{\rm g} - M_\star$ - SFR relation}

\begin{figure}
\begin{center}
\resizebox{8.5cm}{!}{\includegraphics{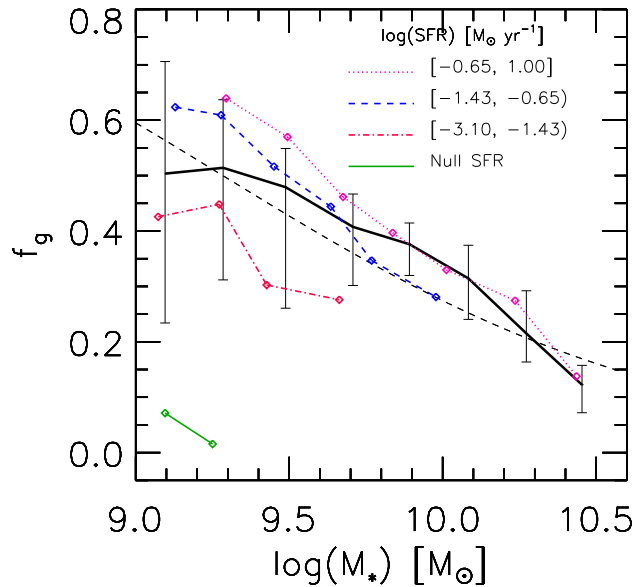}}
\end{center}
\caption[fg vs Ms at $z=0$]
{
Median warm gas fraction, $f_{\rm g}\equiv M_{\rm g}(T<15000~{\rm K})/M_{\rm b}$, as a function of $M_\star$ at redshift $z = 0$, $M_{\rm b}$ is the galaxy's baryon mass ({\em black solid line}), with error bars indicating the 25th and 75th percentiles. The trend is in good agreement with the observed fraction from \cite{stewart2009} ({\em dashed black line}). Coloured lines show the median of $f_{\rm g}$ in the simulation for bins in star formation rate, see legend. At given stellar mass, galaxies with higher gas fractions have higher SFR.}
\label{fig:fg_vs_Ms_z0}
\end{figure}

The anti-correlation of the O/H ratio with gas fraction ($f_{\rm g}$) and SFR at given galaxy stellar mass can be examined in more detail by analysing the  $f_{\rm g} - M_* - {\rm SFR}$ relationship in \gimic. To compare with observations we follow
\citet{derossi2013} and select \lq warm\rq\ gas with temperature\footnote{We reiterate that \gimic\ imposes a pressure-density relation for dense gas, and hence does not distinguish between the warm and cold neutral medium in the ISM of a galaxy.} $T \le 15000$~K, but note that the trends we find remain almost the same if we use the total gas component. 
For $\sim 85\%$ of our sample, this \lq warm\rq\ gas component represents more than $\sim 75\%$ of the total gas content. Gas with $T>15000$~K dominates in mass in only $\sim 7\%$ of \gimic\ galaxies.

The fraction of warm gas in the simulations falls with increasing stellar mass, and follows the observed relation of \cite{stewart2009} very closely (Fig. \ref{fig:fg_vs_Ms_z0}). At given $M_\star$, the simulated $f_{\rm g}$ increases with increasing star formation rate, and galaxies with little or no star formation have very little warm gas. \cite{santini2014} reported the existence of a fundamental $f_{\rm g} - M_* - {\rm SFR}$ relation in observed galaxies, and the \gimic\ galaxies appear to be consistent with this. 
Nevertheless, in the SFR bin where observations and simulations are both available, \gimic\ galaxies tend to over predict gas fractions for massive galaxies. We discuss {\em stellar metallicities} next.

\subsection{Stellar metallicities}
\label{sec:stellar_MZR}
\begin{figure*}
\begin{center}
\resizebox{13cm}{!}{\includegraphics{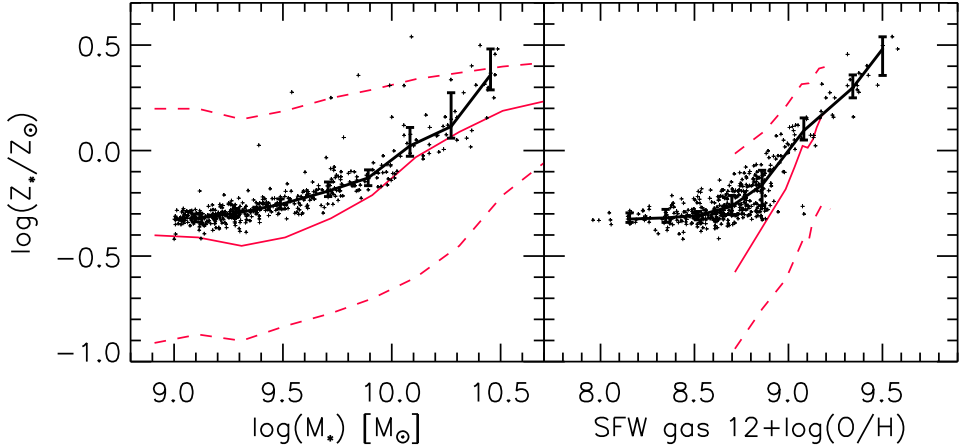}}
\end{center}
\caption[Zs vs Ms and OHg]
{
Median {\em stellar} metallicity as function of stellar mass ({\em left panel}), 
and as function of SFR weighted {\em gas} O/H ({\em right} panel). 
\gimic\ metallicities are shown as symbols, with black line showing the median relation, and the 25th and 75th percentiles indicated as error bars.
Observational data from Gallazzi et al. (2005) are shown as full red lines (median), with dashed lines encompassing the 25th and 75th percentiles. The simulated relation follows the observed relation well in the left panel, but the most massive simulated galaxies are more metal rich as noted before; the scatter in the data is larger than in the simulation. The observed relation in the right panel is steeper than in the simulation, with simulated galaxies reaching a plateau of $\log(Z_\star / Z_{\odot})\sim -0.4$ at low gas abundance, which is not apparent in the data.}
\label{fig:Zs_vs_Ms_OHg}
\end{figure*}

The dependence of stellar metallicity on stellar mass in \gimic\ follows the data from \cite{gallazzi2005} very closely (Fig.~\ref{fig:Zs_vs_Ms_OHg}, {\em left panel}), except for the most massive galaxies ($M_\star\ga 10^{10.5}M_\odot$)
which are overcooled in the simulation 
\footnote{Following \citet{wiersma2009b}, for the solar abundance, we use the metal mass fraction $Z_{\odot} = 0.0127$.}. 
Also, the scatter around the median is significantly larger in the observations. 

We have verified by measuring \gimic\ stellar abundances in apertures ({\em i.e.} varying the value of the parameter $R_{\rm cut}$ discussed earlier), that the smaller scatter in \gimic\ is not likely a consequence of aperture limits. However uncertainties in observed measurements of $Z_\star$ may cause the intrinsic scatter to be overestimated \citep[see][]{gallazzi2005}. Including satellites in our sample {\em does} increase the scatter in the $M_\star-Z_\star$ relation, and we also find that aperture effects {\em do} affect satellites' metallicities. It may be that observational uncertainties, and the inclusion of satellites, would be enough to reconcile observe and simulated scatter, but we have not verified this in detail.

In Fig.~\ref{fig:Zs_vs_Ms_OHg} ({\em right panel}) we explore the relation between {\em stellar} metallicity $Z_\star$ and {\em gas-phase} O/H weighted by star formation rate. We find reasonable agreement with the data of \cite{gallazzi2005}, with the possible exception of the behaviour at low abundance: the simulated $Z_\star$ reaches a plateau in $Z_\star$ at low gas-phase O/H not apparent in the data. This trend in the simulations is consistent with a SF triggered by infall of metal poor gas: in Fig. \ref{fig:OH_vs_Ms_z0} we saw that most metal-poor systems have the highest gas fractions, $f_{\rm g}$, and correspondingly higher star formation rates.

\section{Metallicity evolution}
\label{sec:evolution} 
\begin{figure*}
\begin{center}
\resizebox{17cm}{!}{\includegraphics{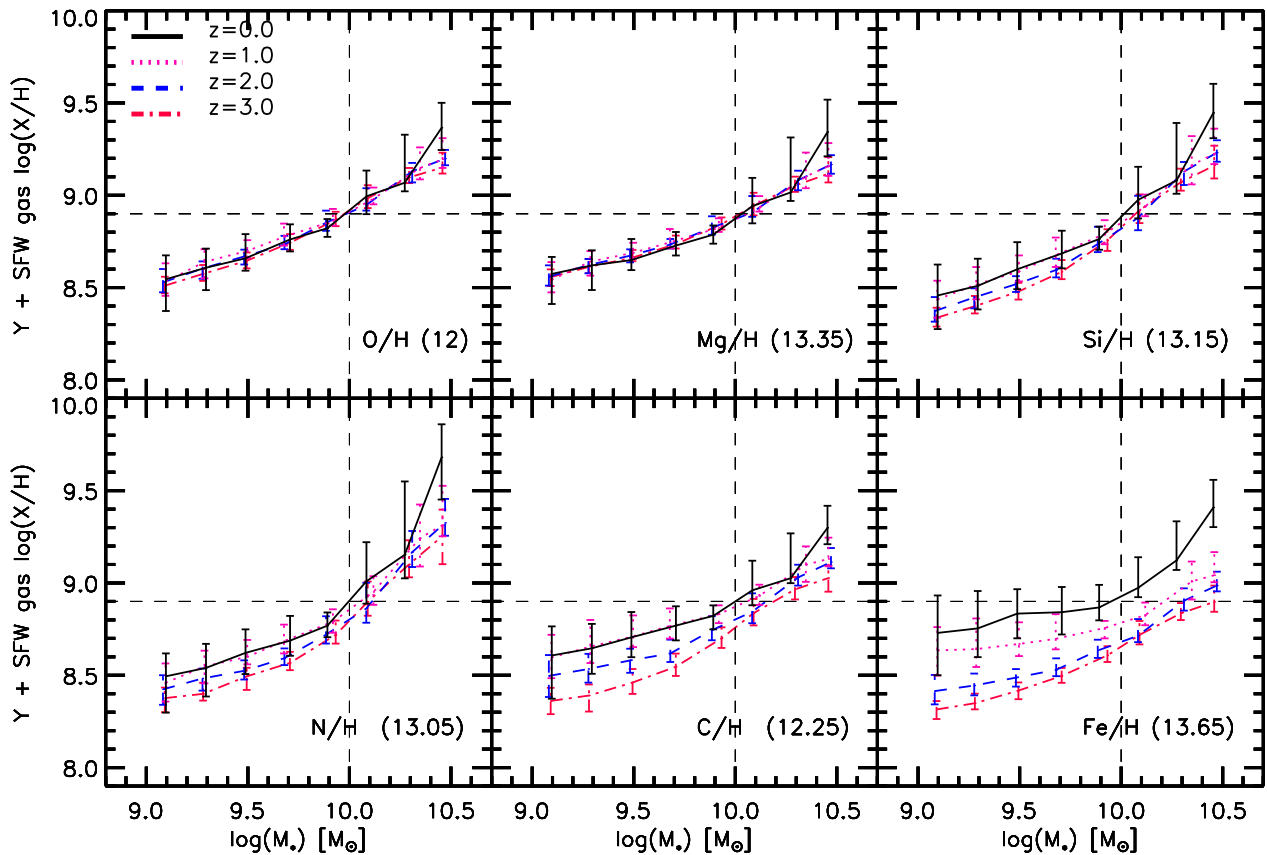}}
\end{center}
\caption[XHg vs Ms]
{
Median SFR-weighted gas abundances as a function of $M_\star$ for different redshifts $z$ (different line styles) for different elements (different panels); error bars depict the 25th and 75th percentiles of galaxies. To ease comparison between different elements, abundances are off-set by a constant - $Y$ - whose value appears in brackets in each panel ($Y=12$ for Oxygen - the usual normalisation for that element). Top panels are for $\alpha$-elements produced in the simulations mainly by type~II SNe, while bottom panels are for N and C, produced by AGB stars, and Fe produced in both type~I and type~II SNe.}
\label{fig: MZR_z}
\end{figure*}

Previously we demonstrated that \gimic\ galaxies display similar trends in ISM and stellar metallicity at redshift $z=0$ as observed galaxies. This reasonable agreement encourages us to explore the abundance enrichment history. In previous sections we have analysed only O/H as Oxygen is the metal most commonly used in observational studies.  In the following, we will take advantage of the enrichment evolution of all elements tracked in the simulation.
We caution the reader that our results of course apply to the particular set of yields discussed in Section \ref{sec:simulation}.

\subsection{Abundance evolution for different elements}
\label{sec:XH_evolution}
The abundance evolution of different elements in the simulation is plotted in Fig.~\ref{fig: MZR_z}, depicting gas phase abundances weighted by star formation rate. The O/H-$M_\star$ relation ({\em upper left panel}) - commonly used as a proxy for the \lq mass-metallicity relation\rq\ does not evolve in \gimic\, either in slope or normalisation, and is in place at least since $z \sim 3$. We verified that this result does not depend on the aperture within which O/H is measured by changing the value of $R_{\rm cut}$ defined earlier. In addition, we employed the redshifts used in the observational analysis of the evolution of the MZR of \citet{maiolino2008}, estimating O/H inside their observed radii: 3.0 kpc (for $z \approx 3.5$), 3.6 kpc ($z \approx 2.25$), 3.725 kpc ($z \approx 0.74$) and 2.0 kpc ($z \approx 0.06$) to enable a fair comparison. Using these apertures has no effect on the slope of the O/H-$M_\star$ relation, but the zero point
increases by 0.2-0.3~dex - a consequence of metallicity gradients in the simulated galaxies. Including satellites in the analysis does introduce modest evolution in O/H with $z$, with higher $z$ galaxies having lower O/H at given stellar mass, see also Section~\ref{sec:satellites} below.

The absence of evolution in the MZR in \gimic\ appears inconsistent with the results of several observational studies which claim {\em significant} evolution since $z=3$, in the sense of higher redshift galaxies having {\em lower} O/H at given stellar mass. For example, \citet{maiolino2008} measure a decrease in O/H for galaxies with $M_\star=10^{10}M_\odot$ of 0.8~dex between redshifts $z=0.7$ and $z=3.5$ (their Figs.~7 and 8), with lower $M_\star$ galaxies evolving even faster; at that stellar mass, $z=2.2$ galaxies are less enriched than $z=0$ galaxies by 0.4~dex.

A plausible explanation for the lack of strong evolution in the MZR in the simulations can be found in \citet{weinmann2012}.  These authors showed that many current cosmological simulations (including \gimic), as well as semi-analytic models, form too much stellar mass too early compared to observations.  As the simulations have approximately the correct amount of stellar mass (per unit halo mass) at $z\sim0.1$, this implies that the simulations will also underpredict the amount of late-time star formation.  A likely consequence of this is that the MZR will be set at higher redshift in the simulations compared to real galaxies.

Alternatively, or perhaps in conjunction with the above, the presence of a strong correlation between O/H-$M_\star$ and $\dot M_\star$ in simulated - and observed - $z\sim 0$ galaxies suggests a possible bias plausibly present in the observations: if higher $z$ galaxies had higher $\dot M_\star$ due to observational selection, then their lower metallicities might be inferred to imply evolution. Indeed, observations by \citet{stott2013} suggest a negligible evolution of the MZR from $z=0$ to $z \sim 1-1.5$, and these authors explain the discrepancy with previous studies due to selection bias. Such an explanation is consistent with \gimic\, in which low-mass galaxies follow the FMR at least up to $z=2$.

Figure~\ref{fig: MZR_z} also displays the evolution of the abundances of other nuclei (${\rm X_i}$).
As expected Oxygen is the dominant metal by mass at all $z$ and as such  is a good proxy for metallicity. The shape of the X$_i$/H-$M_\star$ relation and its evolution (or rather lack thereof) is almost identical for Oxygen and Magnesium, both of which are $\alpha$-elements synthesised in massive stars and released by Type~II SNe. However, the Si/H relation appears to flatten toward $z=0$, even though Si is yet another $\alpha$-element. In the yield tables that \gimic\ uses for Type~II enrichment (basically the yields from \citealt{portinari1998} with some modifications suggested by L. Portinari in a private communication, see \citealt{wiersma2009b} for more details), the metallicity dependence of yields is similar for O and Mg: at higher metallicity, massive stars ($M>30M_\odot$) produce {\em less} O and Mg than at lower metallicity - whereas the behaviour for Si is the opposite ({\em i.e.} at higher metallicity stars tend produce {\em more} Si than at lower metallicity). However in all, there is little if any evolution in the mass-metallicity relation for any of the $\alpha$ elements O, Mg, and Si, with scatter in abundance significantly larger than net evolutionary trends.

The bottom left panels of Fig.~\ref{fig: MZR_z} show evolution of Nitrogen and Carbon. In our models,
AGB stars contribute significantly to C, and they are the dominant source of N. Abundances of N/H and C/H show significant evolution particularly from $z=3\rightarrow 1$, with an increase in abundance towards lower $z$, accompanied by a flattening of the C/H-$M_\star$ relation . In the nucleosynthesis models we employ, the N yield is more sensitive to metallicity than the C yield at higher abundance. This is due to the secondary element nature of N - synthesised from other metals rather just from burning Hydrogen. This might cause the N abundance turn-up for more massive - and hence higher $Z$ galaxies, with approximately N/H proportional to $Z^2$. 
Recent changes to Nitrogen yields at low metallicity discussed by 
e.g. \citet[][]{pipino2009} could plausibly change the evolution of N and a more detailed comparison of the present models to data could test such a model. This is however, beyond the scope of this paper.

The strongest evolution is for Iron ({\em bottom right} panel of Fig.~\ref{fig: MZR_z}), and happens mostly below $z=2$.
Iron has a significant contribution from Type~I SNe which release synthesised Iron with significant delay of $\sim 1$~Gyr after the formation of the progenitor star. At $M_\star=10^{10}M_\odot$, the Fe abundance increases by almost 0.3~dex between $z=3$ and $z=0$, and the increase is even more pronounced at lower $M_\star$.

\begin{figure}
\begin{center}
\resizebox{8.5cm}{!}{\includegraphics{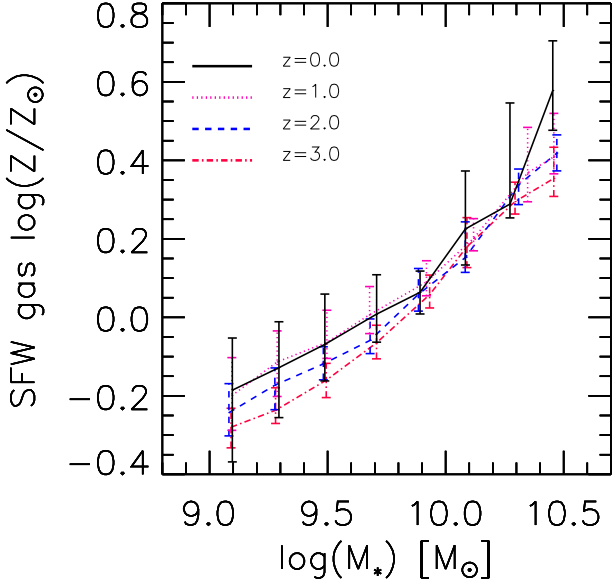}}
\end{center}
\caption[MZR at $z>0$]
{
Median SFR-weighted gas-phase metallicity as a function of $M_\star$ for different redshifts; error bars depict the 25th and 75th percentiles. At a given stellar mass, the ISM is higher enriched at lower $z$ although the trend is weak given the scatter, and is nearly absent for $M_\star\ga 10^{10}M_\odot$.
}
\label{fig:Zg_vs_Ms_z}
\end{figure}

Of course these abundances are not just related to nucleosynthetic yields and how they depend on $Z$, but also to the ages of the population, and to what fraction of the elements synthesised in stars manage to remain in the ISM. \gimic\ galaxies have on average similar mass-weighted stellar ages of $\sim 10~{\rm Gyr}$, therefore comparing galaxies of a given stellar mass, those at lower $z$ had more time to become enriched with elements synthesised in SNIa and AGB stars, driving the observed evolution of non-$\alpha$ elements. The (unrealistically) high ages of simulated galaxies are a consequence of too early star formation in these systems, and leave them with a significant old stellar component. The absence of a $M_\star$-age relation in the simulation might artificially enhance the correlations between abundances and redshift, and may also lead to an underestimate of the scatter in simulated abundances.

The metallicity evolution of the gas phase, weighted by star formation rate, shows only very mild evolution, with $Z$ increasing with time (Fig. \ref{fig:Zg_vs_Ms_z}), with a scatter between galaxies of given $M_\star$ at any given time as large as the evolution of the median $Z$ from $z=3$ to $z=0$. Similarly to the case of stellar abundances this is a consequence of the absence of significant evolution of O/H, and other elements may well show evolution. The different behaviour of Oxygen compared to several other elements suggest that caution should be exercised when using this element just by itself as a measure of abundance evolution.

\subsection{Metallicity as a function of circular velocity}
\label{sec:Z_M_Vcirc}
\begin{figure}
\begin{center}
\resizebox{8.5cm}{!}{\includegraphics{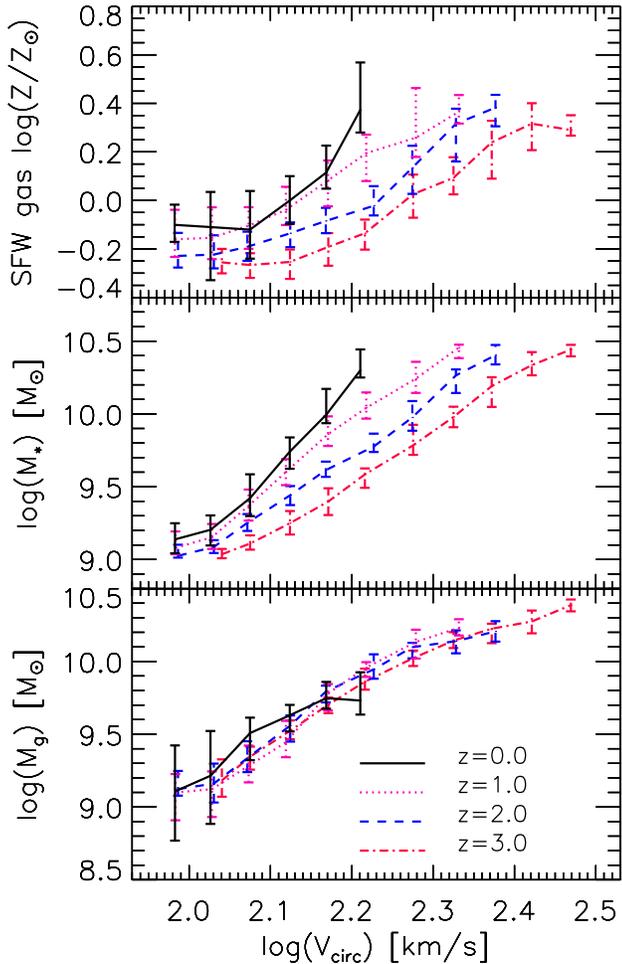}}
\end{center}
\caption[MZR at $z>0$]
{
SFR-weighted gas-phase metallicity (upper panel), stellar mass (middle panel) and
gas mass (lower panel) as a function $V_{\rm circ}$ at various redshifts indicated in the legend.   
Curves denote median relations, with error bars depicting the 25th and 75th percentiles.
At a given $V_{\rm circ}$, galaxies tend to have higher metallicities and higher stellar masses, at lower 
$z$, but similar gas masses. 
}
\label{fig:Z_Ms_Mg_vs_Vcirc_z}
\end{figure}

One theory of the origin of the MZR is that systems with lower $M_\star$ are more affected by galactic winds \citep[e.g.][]{finlator2008}. Such \lq leaky\rq\ galaxies cannot efficiently retain their ISM gas or recently produced metals,
with SNe ejecting both gas and metals from the disk or indeed the halo of such galaxies \citep[see discussion in][]{creasey2015}. Such efficient feedback is a crucial ingredient in models of galaxy evolution \citep[e.g.,][]{schaye2010}, and \gimic\ includes a sub-grid model for launching winds. The extent to which gas mass is ejected from a galaxy by winds is thus more likely to depend on the circular velocity of the galaxy - a measure of the depth of the potential well - rather than its stellar mass, although the gas density and hence the fraction of energy in the wind lost through radiative cooling may play a role as well \citep{creasey2015}. 

We plot the dependence of gas phase metallicity (weighted by star formation rate), stellar mass, and gas mass, as function of circular velocity in Fig. \ref{fig:Z_Ms_Mg_vs_Vcirc_z} for various redshifts.  As expected, systems residing in shallower potential wells exhibit lower metallicities at all $z$, which is consistent with these galaxies being less able to retain metals in the presence of the kinetic energy injected by SNe in our model (see also Section \ref{sec:yields}).
Interestingly the gas mass at a given circular velocity, does not evolve with $z$. As a consequence, galaxies with given circular velocity have significantly higher specific star formation rates, and significantly lower $Z$, with increasing redshift.

We have seen that the value of $Z$ at a given stellar mass does not depend strongly on redshift.
On the other hand, at a given $V_{\rm circ}$, galaxies tend to have significantly 
higher metallicities at lower $z$. The evolution of the $Z$-$V_{\rm circ}$ relation
ranges from $\sim 0.3$ dex at low velocities to $0.6$ dex or more at the high-velocity
end. At first order, the evolution of the $Z$-$V_{\rm circ}$ relation can be understood as a consequence
of the absence of significant evolution of the $Z$-$M_*$ relation but the significant evolution
of the $M_*$-$V_{\rm circ}$  relation (middle panel; see also \citealt{swinbank2012}).
Galaxies with similar $V_{\rm circ}$ (i.e. similarly deep potential wells) have
a more dominant stellar component towards $z \sim 0$ and as a consequence of the MZR, 
higher metallicities.  Because of the more efficient gas outflows of metals at low masses,
the star formation in these systems is from less-chemical enriched material on average.
We also showed that at a high $V_{\rm circ}$, the evolution of $M_*$ is stronger
than in shallower potential wells.  This issue can be probably related to the increase of the 
merger events for larger galaxies which do not only add the mass of the merging
satellites but also contribute to the enhancement of the SF rate.

In the lower panel of Fig. \ref{fig:Z_Ms_Mg_vs_Vcirc_z}, 
we show the gas-phase mass in simulated galaxies as a function
of $V_{\rm circ}$; interestingly there is almost negligible 
evolution. The increase of $M_*$ at a fixed potential well, with
no significant variations of the gas mass, implies that 
the gas fractions of simulated galaxies decrease towards lower redshifts, consistent
with the observed behaviour. This means that if the potential well
of an individual galaxy does not change significantly over a given
period of time and if this galaxy does not experience mergers over that period, 
its metallicity will increase as gas is converted into stars which enrich 
the ISM. In addition, as star formation proceeds consuming the gas-phase, 
gas infall will be required to maintain an almost constant gas component.

\subsection{Abundance imprints of galactic outflows}
\label{sec:yields}

\begin{figure*}
\begin{center}
\resizebox{13cm}{!}{\includegraphics{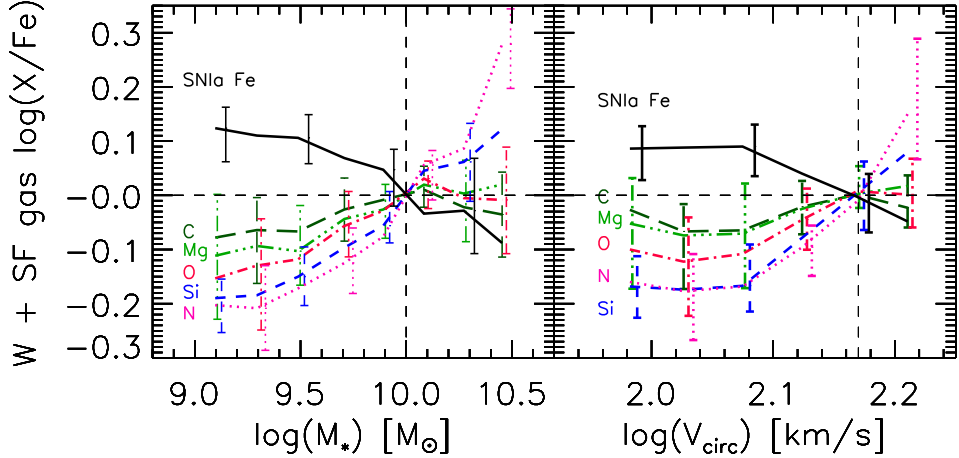}}
\end{center}
\caption[SFgasXFe vs Ms and Vcirc]
{Median abundance of X/Fe for various elements X (X=N, Si, O, Mg, C and Fe from type Ia SNe only, as indicated in the legend), in star forming gas at redshift $z=0$ as function of stellar mass $M_\star$ and circular velocity $V_{\rm circ}$ ({\em left} and {\em right} panels, respectively); error bars denote the 25th and 75th percentiles. For the sake of clarity, not all error bars are shown and 
we applied a small offset in the horizontal direction  between some error bars
to improve visibility. Abundances are off-set vertically by a constant $W$ chosen such that they are all equal at $M_\star=10^{10} M_{\odot}$ ($V_{\rm circ}=150$~km$^{-1}$ in the right panel), $W=$  -0.685 (C), 0.127 (Mg), -1.08 (0), -0.14 (Si), 0.03 (N) and 0.36 (Fe from SNIa only). The ratio N/Fe increases dramatically with increasing $M_\star$ or $V_{\rm circ}$, whereas the contribution of Fe from type~I SNe decreases; these two elements bracket the behaviour of the others.}
\label{fig:XFe_vs_Ms_Vcirc}
\end{figure*}

As we have seen, our simulations are consistent with a scenario in which galactic outflows play an important role in setting the level of depletion of metals. We further investigate this process here by contrasting the abundances of elements produced by different channels, namely SNII, SNIa and AGB stars. Outflows in \gimic\ are driven by SNe explosions, which are assumed to occur $\sim 30$~Myr after star formation. Every star particle that is 30~Myr old imparts kinetic energy in the form of a randomly directed kick with a launch velocity of 600~km~s$^{-1}$ to, on average, $\eta=4$ of its neighbouring gas particles. Type~II, type~I and AGB nucleo-synthetic produce are donated to neighbouring gas particles during each time-step. Since type~II produce is distributed to neighbouring particles just before they are kicked, these elements are more likely to escape a galaxy in a galactic wind than elements produced by type~I SNe and AGB stars, whose metals are released from much older stars and after the initial SNII kick.

\cite{wiersma2010} discusses the enrichment of cosmic metals in the {\sc OWLS} suite, which used the same code, has comparable resolution to \gimic\ for the high resolution runs, but were performed in periodic boxes. These authors find that metals residing in lower density gas were typically ejected earlier and by winds driven from lower mass galaxies. \cite{wiersma2011} investigate to what extent metal enrichment depends on the relatively poorly constrained sub-grid modelling of feedback. Their Fig.~12 shows that in most models stars are the dominant repository of metals at redshift $z=0$ ($\sim 60-70$ per cent for the {\sc OWLS} reference and {\sc MILL} models for which the feedback implementation is most similar to that of \gimic). However in the AGN model that includes feedback from accretion black holes, this decreases to 20 per cent. The second main repository of metals in these models is the warm-hot intergalactic medium (WHIM). These results demonstrate that the metallicity of stars or star forming gas in \gimic\ is not just set by the star formation efficiency or yields: metal mass loss resulting from feedback indeed plays an important role. For halos of 
mass below $\sim 10^{12}M_\odot$ feedback is regulated mostly by
SNe driven winds, whereas in more massive haloes AGN feedback may be more important. Note that \gimic\ does not include AGN.

We plot the SF gas abundance X/Fe (X=N, Si, O, Mg, C and Fe from type~I SNe only) as a function of $M_\star$ and as function  of $V_{\rm circ}$ at $z = 0$ in Fig. \ref{fig:XFe_vs_Ms_Vcirc} ({\em left} and {\em right} panel, respectively); abundances are off-set vertically so that the lines cross at $M_\star=10^{10}M_\odot$ ($V_{\rm circ}=150$~km~s$^{-1}$ for the right panel).
The ratio $\alpha$/Fe of $\alpha$ elements ($\alpha$=O, Si, Mg which are produced in type~II SNe) compared to (total) Fe increases with increasing mass up to $M\sim 10^{10}M_\odot$, then flattens for O and Mg but keeps rising for Si; a similar trend is seen in $\alpha/$Fe as function of $V_{\rm circ}$. This ratio depends on the star formation time scale, the delay time distribution of type~I SNe, and whether or not type~II produce is more likely to escape the galaxy in a wind than type~I produce.  However it of course also depends on nucleo-synthetic yields, which themselves depend on metallicity and hence stellar mass - this is the reason these elements do not track each other in detail in \gimic\ . The X/Fe-$M_\star$ relations for the stellar component are similar to those of star forming gas - modulo minor changes in the zero point.

\begin{figure*}
\begin{center}
\resizebox{13cm}{!}{\includegraphics{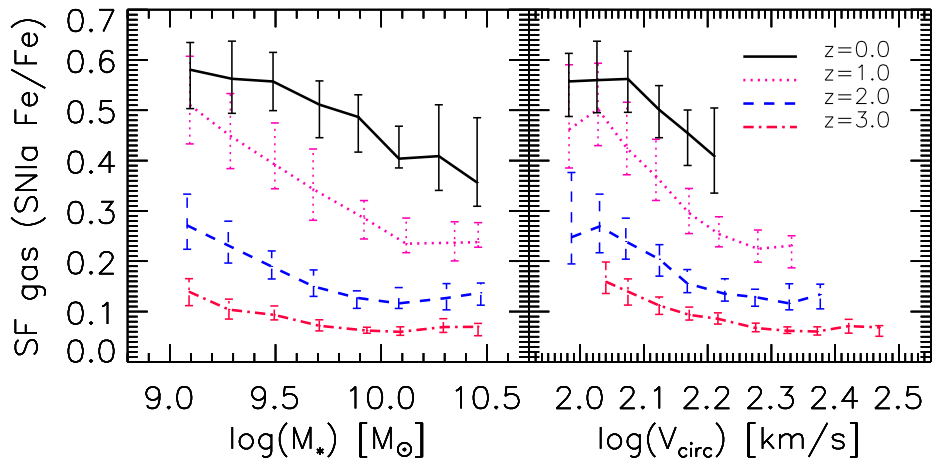}}
\end{center}
\caption[f_FeSNIa vs Ms and  Vcirc]
{Median mass fraction of Fe from SNIa in star forming gas as a function of $M_\star$ ({\em left panel}) and $V_{\rm circ}$ ({\em right panel}) for various redshifts (different curves); error bars denote 25th and 75th percentiles.
The contribution to total Fe from type~I SNe increases with time as expected from the delayed production of type~I Fe in the model. The dependence with $M_\star$ (or $V_{\rm circ}$) due to more significant loss of SNII produce from shallower potentials, also becomes more pronounce at later times.}
\label{fig:FeSNIa_vs_Ms_Vcirc}
\end{figure*}

We mentioned in Section \ref{sec:XH_evolution} that simulated galaxies have 
around the same mass-weighted stellar age ($\sim 10 {\rm Gyr}$).  
Given the similar mean ages of our galaxies, 
the slope of the X/Fe-$M_*$ relation will be significantly influenced by 
the effects of galactic winds at different masses.
From Fig. \ref{fig:XFe_vs_Ms_Vcirc}, left panel, 
it is clear that there is a correlation between $M_*$ and X/Fe for C, Mg, O, Si
and N and an anti-correlation between $M_*$ and the fraction of Fe from SNIa ($f_{\rm Fe,SNIa}$).  
These results are consistent with what we would expect if low-mass systems were more affected
by galactic outflows. The shallower potential wells where low-mass systems reside are
less efficient in retaining metals against galactic winds generated by SNII.  On
the other hand, due to the longer time scales for the release of SNIa yields, 
Fe from SNIa is not
directly affected by these winds, generating lower X/Fe.  It is worth noting that
C and N are mainly generated by AGB stars but the contribution from SNII is also significant 
\citep{wiersma2009b}
so that they are also affected by winds. In the right panel of Fig. \ref{fig:XFe_vs_Ms_Vcirc},
we can see the dependence of X/Fe on the potential well of simulated galaxies.
Median X/Fe ($f_{\rm Fe,SNIa}$) seem to be constant at $\log (V_{\rm circ}) < 2.1$
and increase (decrease) for larger $\log (V_{\rm circ})$, indicating that galactic
winds dominates at low $V_{\rm circ}$ values, as expected.

We note that the slope of the  X/Fe-$M_*$ relation is shallower for C/Fe, 
probably caused by the significant  contribution of AGB stars to C enrichment.
A steeper slope is obtained for Si/Fe and N/Fe. In the case of Si, this may be related to the assumed yields 
of \citet{portinari1998} for which Si production decreases with the metallicity of the progenitor
star.

The interpretation of the trends is more convoluted for the case of N which has characteristics of both a primary and a secondary elements. 
At low $Z$, N production results mainly from burning the C and O generated during the star's evolution.
Consequently its yield tracks the yields of C and O, and N behaves as a primary element 
However at higher $Z$,
N production results increasingly from burning the C and O that was already in the star from birth and N behaves more as a secondary element - the N yield no longer tracks the C/O yield but depends on $Z$.  Therefore, for galaxies with low metallicity, N is expected to behave as a primary element showing no important variations in N/O while for more metal-enriched galaxies, N/O should increase with O/H as N behaves as a secondary element.
Because of the tightness of the MZR (Fig.\ref{fig:Zs_vs_Ms_OHg}), the abundance of O/H is almost a proxy for $M_*$. This reasoning explains why the dependence of N abundance on $M_\star$ steepens increasingly with increasing $M_\star$, as seen in Fig. \ref{fig:XFe_vs_Ms_Vcirc}. The behaviour of N/O with stellar mass may be different if
Nitrogen were produced abundantly in low-Z fast-rotating stars, as in the models of \cite{pipino2009}.

The study of the stellar X/Fe as a function of $M_*$  for elliptical galaxies is discussed extensively in the literature \citep[see e.g., ][]{conroy2013}. Although a detailed analysis is beyond the scope of the present paper, we did perform a preliminary analysis of the stellar X/Fe-$M_*$ relation using 
a subsample of galaxies with large velocity dispersion, typical of elliptical systems.  We find larger average values of [Mg/Fe], [Si/Fe] and [N/Fe] than reported by \citet{conroy2013} by $\sim 0.1$ dex, $\sim 0.3$ dex and $\sim 0.15$ dex, respectively. Observational uncertainties are of order 0.05~dex \citep{conroy2013} and differences with other observational works are of order $\sim 0.1$ dex. Uncertainties in yields, and in the rates of SNIa, may be enough to explain the remaining differences.

Finally we compare the abundance evolution of yields from SNII and SNIa as a function of $M_*$ and $V_{\rm circ}$ (Fig. \ref{fig:FeSNIa_vs_Ms_Vcirc}), by plotting the fraction of Fe produced by SNIa. We see a significant trend of increasing SNIa Fe/total Fe with decreasing z, and decreasing $M_\star$ (or $V_{\rm circ}$). The evolution with $z$ is stronger for lower mass galaxies. These trends are consistent with our findings that type II products escape more easily from lower mass galaxies, and with the delay of SNIa enrichment. We reiterate that we may be overestimating the SNIa contributions especially for lower mass galaxies, because the stars in our low mass galaxies tend to be too old on average.

\section{Satellite galaxies}
\label{sec:satellites}
\begin{figure}
\begin{center}
\resizebox{7cm}{!}{\includegraphics{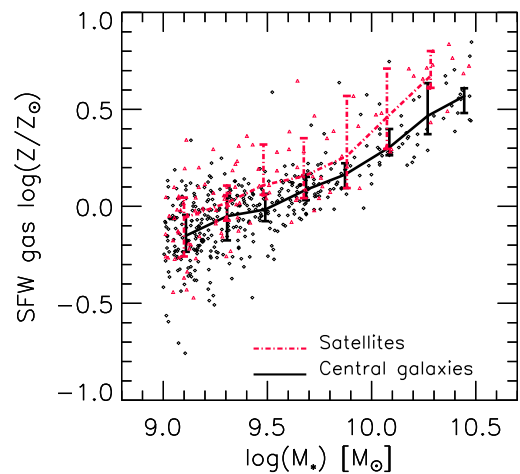}}
\end{center}
\caption[MZR for central and satellite galaxies]
{
SFR-weighted gas-phase metallicity as a function of stellar mass at $z=0$ for
central ({\em black}) and satellite ({\em red}) galaxies; curves 
depict the median relations with error bars indicating the 25th and 75th percentiles. 
At a given $M_*$, satellites have higher median metallicities by $\sim 0.1-0.2$~dex, and their scatter
in $Z$ is also significantly higher. Both contribute to making the scatter in Z at given $M_\star$ much higher for a sample of galaxies that includes both central and satellites compared to a sample consisting of just centrals.}
\label{fig:MZR_sat_z0}
\end{figure}
\begin{figure}
\begin{center}
\resizebox{7cm}{!}{\includegraphics{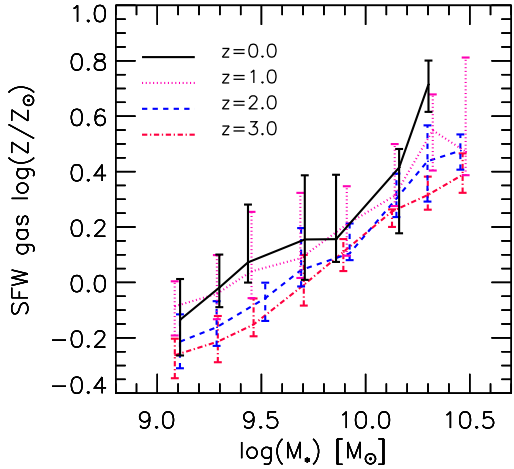}}
\end{center}
\caption[MZR for satellite galaxies at $z \ge 0$]
{
Median SFR-weighted gas-phase metallicity $Z$ as a function of stellar mass for
satellite galaxies at different redshifts $z$; error bars depict the 25th and 75th percentiles. Metallicity increases with $M_\star$, but unlike the case of centrals plotted in Fig.~\ref{fig: MZR_z}, $Z$ also evolves with $z$, with satellites $\sim 0.2-0.4$~dex more metal rich at $z=0$ compared to $z=3$ at given stellar mass. The scatter in $Z$ at given $M_\star$ is also significantly larger than for centrals.}
\label{fig:MZR_sat_z}
\end{figure}

In Section~\ref{sec:simulation} we defined a central galaxy as the most massive {\sc subfind} structure in a given friends-of-friends halo -- all other galaxies in that halo are satellites. Here we investigate abundance trends in satellites using the same criteria as for centrals regarding number of particles and mass range (see Section \ref{sec:sample}). Consequently our sample will only include the most massive satellites;
typically only $20-30$\% of the total depending on redshift.

The MZR of satellites and central galaxies at $z=0$ are compared in Fig.~\ref{fig:MZR_sat_z0}.
At given $M_*$, satellites are on average more metal rich than centrals by  $\sim 0.1-0.2$ dex, and their metallicity scatter is higher than for centrals ($\sim 0.5$~dex for satellites compared to $\sim 0.1-0.2$~dex for centrals), mainly at intermediate masses $M_\star\sim 10^{10}M_\odot$.  As we noted in Section \ref{sec:stellar_MZR}, the inclusion of satellites in the analysis of the MZR based on stellar abundances 
tends to increase the scatter.  Our findings indicate that this is also true for gas abundances; their inclusion also results in an offset to higher $Z$.

\citet{pasquali2012} compare the MZR of satellite and central galaxies inferred from SDSS data, finding that satellites tend to have higher metallicities than central systems at a given $M_*$ by $\sim 0.06$ dex, with the offset increasing with decreasing $M_\star$. We find a slightly larger offset, and no obvious trend with $M_\star$ except at the very massive end. Previously we explained why varying the size of the aperture within which $Z$ is measured affects the zero point of the MZR because of abundance gradients - we find the same to be true for satellites. It is possible that this affects the interpretation of the data from SDSS due to the size of the fibres. 

What is the origin of the differences in Z between satellites and centrals? 
At a given $M_*$, satellites reside in shallower potential wells than central galaxies as
expected: their $V_{\rm circ}$ are lower by $\sim 0.05$~dex than those of central
galaxies of similar masses. Moreover, the $f_{\rm g}$-$M_*$ relation of satellite galaxies is offset by -0.2 when 
compared to centrals, mainly due to the loss of non star-forming gas. This lower gas fraction of satellites is consistent with these systems being subject to strangulation and ram-pressure stripping. We also find that the scatter in $f_{\rm g}$-$M_*$ (and $f_{\rm g}-V_{\rm circ}$) is larger for satellites than for centrals. These findings are consistent with a scenario in which satellite galaxies
loose the metal poor gas in their outskirts due to ram-pressure stripping. This gas is then no longer available to dilute the ISM of the satellite, which leads to them having a higher $Z$. The larger scatter in $f_{\rm g}$ is also consistent with the larger scatter in $Z$ at given $M_\star$ in satellites compared to centrals. Such a scenario can potentially explain both the offset and increase in scatter in $Z$, but we postpone a more detailed investigation to future work.

The evolution of the MZR for satellites is shown in Fig.~\ref{fig:MZR_sat_z}, which can be compared to the 
evolution for centrals in Fig.~\ref{fig: MZR_z}. We recall that the ratio O/H - a good proxy for $Z$ - does not show any appreciable evolution for centrals. In contrast $Z$ does evolve for satellites, increasing from $z=3$ to $z=0$ by $\sim 0.2-0.4$~dex, depending on $M_\star$. Interestingly at $z \sim 3$, satellites and centrals have MZRs with similar amplitude and slope.  We have verified that simulated satellites inhabit halos typically 10 times more massive than centrals at a given $M_*$, which contributes to the higher metallicity of satellites.

\section{Discussion}
\label{sec:discussion}
Galaxies in the {\sc gimic} simulation follow a mass-metallicity relation consistent with that observed, although the evolution of the MZR may be less than that observed. We argued this maybe at least partly a consequence of bias in the observations. As shown in Sections~ \ref{sec:Z_M_Vcirc} and  \ref{sec:yields}, the slope of the MZR is to first order set by the efficiency with which galactic winds can escape the potential well. In particular type II SN products such as Oxygen escape more readily from shallower potential wells. Since Oxygen is often used as a proxy for metallicity $Z$, \lq metals\rq\ also escape more easily from shallower potential wells. We note that this does not immediately imply that it is the escape velocity that determines whether metal-enriched winds can escape from a galaxy: higher cooling loses in a denser ISM may play an important role as well \citep{creasey2015}. We also noted in Section~\ref{sec:local_mzr} that the simulated MZR does not exhibit the observed  flattening at the high mass end, which we counted on the absence of AGN feedback in {\sc gimic}. We conclude that feedback from SNe modulates the slope of the MZR at low and intermediate masses while the flattening at higher masses is increasingly affected by AGN.

While galactic winds appear to be the main process for setting the slope of the MZR, its zero
point is also affected by the SF time scales. The zero point of the X/H-$M_*$ relation for the $\alpha$ elements 
produced by short-lived stars does not exhibit significant evolution since $z \sim 3$. We may have underestimated
the level of evolution in $\alpha$ elements because our galaxies tend to be too old.  However, the ratio X/H-$M_*$ at given $M_*$ does evolve for elements produced by SNIs and AGB stars, and the level of enrichment is connected to the SF time scale.

Applying observed aperture corrections and including satellites in our analysis increase the level of evolution by up to $\sim 0.4$~dex, however it is still not as strong as many observers infer.  We believe this is at least partly due to the fact that stellar populations in the simulation are typically older than in the real universe, plausibly because feedback in the simulation is not sufficiently effective at early times
\citep[see also][]{weinmann2012,derossi2013}. In addition we find that typical specific star formation rates in {\sc gimic} 
are too low at redshift $z=0$, in particular for low-mass galaxies, as also found by \citet{mccarthy2012} who blamed this partly on lack of numerical resolution.  We note, however, that the overall MZR is not strongly affected by resolution, as demonstrated in the Appendix. 
In the recent {\sc eagle} simulations \citep{schaye2015}, feedback becomes more efficient at higher density which suppresses star formation at early times and increases the specific star formation at later times. This may help to increase the level of evolution in the simulations.

With respect to the scatter in the MZR, our findings indicate that at a given $M_*$, metallicities show a secondary dependence on $f_{\rm g}$, where galaxies with lower metallicities tend to have higher $f_{\rm g}$ and higher SFRs,  consistent with the observed trends.  Therefore, the SF process in these systems
is driven mainly by accretion of metal poor gas. 

Interestingly, we find that, at a given $M_\star$, systems with lower gas-phase metallicities tend to exhibit {\em higher} $V_{\rm circ}$. Therefore, at a given $M_\star$, systems with deeper potential wells have larger fractions of gas which is also more metal-poor. Such a correlation suggests that the depth of the potential well at a fixed $M_\star$ is regulating the infall of 
metal-poor gas 
and hence, the scatter in the MZR.  In addition, at a fixed $M_*$, the radius which encloses 50\% of the stellar component ($R_{50}$) tends to be larger for systems with lower metallicities, evidencing again the more significant influence of gas infall in these galaxies.

In summary: the zero point of the MZR in the {\sc gimic} simulations is at first order set by the star formation time scale, and the efficiency of galactic winds, respectively. The scatter around the relation is affected by the gas fraction, itself correlated with the depth of the potential of a galaxy of given mass.
The MZR of satellites is off-set to higher $Z$ compared to centrals of the same mass, and its scatter and level of evolution are significantly higher.  We verified that ram-pressure stripping and the metal pollution of the host halos in which satellites live, seem to play a key role on modulating their metal enrichment.

\section{Conclusions}
\label{sec:conclusions}

We have analysed the {\sc gimic} hydrodynamical simulations of structure formation to study the abundances
of different elements in star forming gas in galaxies as a function of galaxy stellar mass and redshift. We focused mainly on the analysis of central substructures of dark matter halos (\lq central galaxies\rq) but we also looked at satellite systems. Our main conclusions can be summarised as follows:

\begin{itemize}

\item {\sc gimic}  galaxies exhibit a well-defined correlation between
O/H and $M_*$ at $z = 0$ in the sense that more massive galaxies are more metal-enriched,
in agreement with observations (Fig.~\ref{fig:OH_vs_Ms_z0}). However, at the high-mass end ($M_\star\ga 10^{10.5} M_\odot$) simulations predict higher abundances than observations.  We suggest that the inclusion of AGN feedback in the model would help to overcome this problem by preventing overcooling in massive galaxies.

\item  Consistent with observations, the scatter of the O/H-$M_*$ relation is driven by 
a secondary dependence of metallicity on the gas fraction ($f_{\rm g}$, Fig.~\ref{fig:fmr}).
At given $M_\star$, galaxies with lower metallicities $Z$ tend to have higher $f_{\rm g}$, higher star formation rates, inhabit deeper potential wells and have larger half-mass radii. Such correlations are expected if star formation in galaxies is self-regulating \citep{schaye2010} with feedback regulating inflow of more metal poor gas.

\item The simulations are able to reproduce the correlation between O/H, $M_\star$ and SFR of the fundamental mass-metallicity relation (FMR) of observed galaxies of \citet{mannucci2010}. Our results are also consistent with the existence of a correlation between O/H, $M_*$ and $f_{\rm gas}$.  Following observational parametrizations  of the FMR, we analysed the parameters
${\mu}_{\alpha}$ and ${\eta}_{\beta}$. We found that $\beta \sim {0.48}$ leads to the tightest relation with O/H, reducing the scatter by $\sim 0.03$ dex with respect to the O/H-$M_*$ relationship. Our findings show also evidence for a correlation between $f_{\rm g}$, $M_*$ and SFR, 
consistent with the observations by \citet{santini2014}.  At a given $M_\star$, simulated systems
with higher $f_{\rm g}$ tend to have higher SFRs. The simulated $f_{\rm g}$-$M_*$ relation for the cold gas is in remarkably good agreement with observations.

\item The $Z_\star$-$M_\star$ relation in {\sc gimic} agrees well with observations in both slope and zero point (Fig.~\ref{fig:Zs_vs_Ms_OHg}); however the scatter in the simulations is lower. Including satellites in the analysis, and taking into account aperture effects increases the scatter in the simulation considerably. 

\item There is negligible evolution of the O/H-$M_*$ relation between $z=3$ and $z=0$ (Fig.~\ref{fig: MZR_z}), in apparent disagreement with observations. Taking into account satellite galaxies - that do evolve - and aperture effects leads combined to a small evolution of $\sim 0.4$~dex from $z=3$ to $z=0$, still considerably less than the claimed $1$~dex evolution in the data. In {\sc gimic}, galaxy stellar ages tend to be uniformly old, $\sim 10$~Gyr which contributes to the disagreement - modifying the stellar feedback efficiency as in the recent {\sc eagle} simulations \citep{schaye2015} might resolve some of the discrepancy.
However we suggest that observational bias in the data may be the main culprit.

\item The evolution of the abundance ratios (X/H) as a function of galaxy stellar mass,
for various other elements X tracked in the simulation, depends strongly on the stellar evolutionary channel 
that produces X (Fig.~\ref{fig: MZR_z}). $\alpha$-elements produced by massive stars show very little evolution, whereas elements with significant SNI or AGB contributions show stronger evolution.  Therefore, the total metallicity ($Z$) of the gas evolves in the sense that at a given mass, systems were less enriched in the past. The different behaviour of these elements can be understood in terms of ({\em i}) the ejection of SNII produce in galactic winds, and ({\em ii}) the stellar evolutionary time scale for SNI and AGB stars. Galactic winds are driven by the same type II SNe that also release $\alpha$ elements, and such metal enriched winds escape easily from small galaxies living in a shallow potential well. If the stars in these galaxies are old enough to produce SNI and AGB stars, then elements produced predominantly by SNI and AGB stars dominate the abundance pattern. More massive galaxies from which winds cannot escape easily are then relatively more abundant in $\alpha$ elements. 
We focused mainly on the comparison of the slope of the different abundance ratios as the theoretical yields are uncertain by a factor of a few.
A comparison between results presented here and those derived from other yields models available in the literature
would be very interesting.

\item Satellites are more metal rich than centrals at given $M_\star$ in agreement with observations, and the scatter around the mean relation is significantly larger for satellites (Fig.~\ref{fig:MZR_sat_z0}). The inclusion of satellites in the analysis of the MZR could
increase the scatter of the relation, leads to an offset, and introduces evolution.  

Satellites have similar amounts of star forming gas, and similar star formation rates as central galaxies
at a given $M_\star$. However, they have less gas in total, probably because of strangulation
and ram-pressure stripping \citep[see also][for a more
extensive study of the satellite galaxy population in \gimic]
{bahe2013, bahe2015}.  Their depletion in gas contributes to their higher metallicities.  

\end{itemize} 

The mass-metallicity relation in {\sc gimic} is mostly determined by the star formation time scales and the efficiency of SN feedback as a function of galaxy stellar mass.  SN feedback is efficient in shaping the relation at low and intermediate masses, $M_\star\la 10^{10.5} M_\odot$, but at higher masses the absence of AGN feedback in the simulation leads to unrealistically massive galaxies. 
The evolution of the zero point of the MZR seems to be directly related to the SF time scales.
Since \gimic\ galaxies have a dominant older stellar component, the MZR based on $\alpha$ elements alone
does not evolve significantly.  We suggest that an improvement of the feedback prescription as implemented in {\sc eagle} \citep{schaye2015} might affect this. At second order, the simulated MZR shows dependencies on the infall of metal-poor gas. At a given $M_\star$, systems with deeper potential wells seem to acrete more metal-poor gas increasing their SFRs. These findings are consistent with previous
proposed scenarios in the literature.

Finally, although there are still important discrepancies between observational studies
with respect to the shape, zero point and evolution of the MZR and its 3D extensions, the advances
of the observational techniques are shedding more light on the 
abundance evolution of galaxies over cosmic time.
The continued comparison between these observations with theoretical models of galaxy formation not only offers a fruitful means for testing/constraining the models, but also helps to guide those
studies and provide plausible formation scenarios.  
A significant combined effort will be required from the theoretical and observational
sides in order to converge to a complete understanding of the abundance
enrichment history of galaxies.

\section*{Acknowledgements}
We thank the referee for constructive remarks which improved the paper.  We also thank Russell Smith, Richard Bower and John Stott for useful comments.
M.E.D.R. is grateful to the ICC staff for their hospitality during her visits and
to Mar\'{\i}a Sanz and Guadalupe De Lucia for their help and support.
We acknowledge the LACEGAL People Network supported by the European Community.
This work was supported by the Science and Technology Facilities Council 
[grant number ST/F001166/1], and by the Interuniversity Attraction Poles 
Programme initiated by the Belgian Science Policy Office ([APP7/08 CHARM].
M.E.D.R. acknowledge support from the PIP 2009-112-200901-00305 of CONICET (Argentina) 
and the PICT Raices 2011-0959 of ANPCyT (Argentina). 
This work used the DiRAC Data Centric system at Durham University, operated by the Institute for Computational Cosmology on behalf of the STFC DiRAC HPC Facility (www.dirac.ac.uk). This equipment was funded by BIS National E-infrastructure capital grant ST/K00042X/1, STFC capital grant ST/H008519/1, and STFC DiRAC Operations grant ST/K003267/1 and Durham University. DiRAC is part of the National E-Infrastructure.

\bibliographystyle{mn2efix.bst}

\bibliography{references}

\begin{thebibliography}{104}
\expandafter\ifx\csname natexlab\endcsname\relax\def\natexlab#1{#1}\fi

\bibitem[{{Altay} {et~al}\mbox{.}(2011){Altay}, {Theuns}, {Schaye}, {Crighton},
  \& {Dalla Vecchia}}]{altay2011}
{Altay} G., {Theuns} T., {Schaye} J., {Crighton} N.~H.~M., {Dalla Vecchia} C.,
  2011, \apjl, 737, L37

\bibitem[{{Andrews} \& {Martini}(2013)}]{andrews2013}
{Andrews} B.~H., {Martini} P., 2013, \apj, 765, 140

\bibitem[{{Bah{\'e}} \& {McCarthy}(2015)}]{bahe2015}
{Bah{\'e}} Y.~M., {McCarthy} I.~G., 2015, \mnras, 447, 969

\bibitem[{{Bah{\'e}} {et~al}\mbox{.}(2013){Bah{\'e}}, {McCarthy}, {Balogh}, \&
  {Font}}]{bahe2013}
{Bah{\'e}} Y.~M., {McCarthy} I.~G., {Balogh} M.~L., {Font} A.~S., 2013, \mnras,
  430, 3017

\bibitem[{{Bothwell} {et~al}\mbox{.}(2013){Bothwell}, {Maiolino}, {Kennicutt},
  {Cresci}, {Mannucci}, {Marconi}, \& {Cicone}}]{bothwell2013}
{Bothwell} M.~S., {Maiolino} R., {Kennicutt} R., {Cresci} G., {Mannucci} F.,
  {Marconi} A., {Cicone} C., 2013, \mnras, 433, 1425

\bibitem[{{Bower} {et~al}\mbox{.}(2006){Bower}, {Benson}, {Malbon}, {Helly},
  {Frenk}, {Baugh}, {Cole}, \& {Lacey}}]{bower2006}
{Bower} R.~G., {Benson} A.~J., {Malbon} R., {Helly} J.~C., {Frenk} C.~S.,
  {Baugh} C.~M., {Cole} S., {Lacey} C.~G., 2006, \mnras, 370, 645

\bibitem[{{Brooks} {et~al}\mbox{.}(2007){Brooks}, {Governato}, {Booth},
  {Willman}, {Gardner}, {Wadsley}, {Stinson}, \& {Quinn}}]{brooks2007}
{Brooks} A.~M., {Governato} F., {Booth} C.~M., {Willman} B., {Gardner} J.~P.,
  {Wadsley} J., {Stinson} G., {Quinn} T., 2007, \apjl, 655, L17

\bibitem[{{Calura} {et~al}\mbox{.}(2009){Calura}, {Pipino}, {Chiappini},
  {Matteucci}, \& {Maiolino}}]{calura2009}
{Calura} F., {Pipino} A., {Chiappini} C., {Matteucci} F., {Maiolino} R., 2009,
  \aap, 504, 373

\bibitem[{{Chabrier}(2003)}]{chabrier2003}
{Chabrier} G., 2003, \apjl, 586, L133

\bibitem[{{Chiappini} {et~al}\mbox{.}(2006){Chiappini}, {Hirschi}, {Meynet},
  {Ekstr{\"o}m}, {Maeder}, \& {Matteucci}}]{Chiappini06}
{Chiappini} C., {Hirschi} R., {Meynet} G., {Ekstr{\"o}m} S., {Maeder} A.,
  {Matteucci} F., 2006, \aap, 449, L27

\bibitem[{{Conroy}, {Graves} \& {van Dokkum}(2014){Conroy}, {Graves}, \& {van
  Dokkum}}]{conroy2013}
{Conroy} C., {Graves} G.~J., {van Dokkum} P.~G., 2014, \apj, 780, 33

\bibitem[{{Cooper} {et~al}\mbox{.}(2008){Cooper}, {Tremonti}, {Newman}, \&
  {Zabludoff}}]{cooper2008}
{Cooper} M.~C., {Tremonti} C.~A., {Newman} J.~A., {Zabludoff} A.~I., 2008,
  \mnras, 390, 245

\bibitem[{{Crain} {et~al}\mbox{.}(2009){Crain}, {Theuns}, {Dalla Vecchia},
  {Eke}, {Frenk}, {Jenkins}, {Kay}, {Peacock}, {Pearce}, {Schaye}, {Springel},
  {Thomas}, {White}, \& {Wiersma}}]{crain2009}
{Crain} R.~A. {et~al.}, 2009, \mnras, 399, 1773

\bibitem[{{Creasey}, {Theuns} \& {Bower}(2015){Creasey}, {Theuns}, \&
  {Bower}}]{creasey2015}
{Creasey} P., {Theuns} T., {Bower} R.~G., 2015, \mnras, 446, 2125

\bibitem[{{Cresci} {et~al}\mbox{.}(2012){Cresci}, {Mannucci}, {Sommariva},
  {Maiolino}, {Marconi}, \& {Brusa}}]{cresci2012}
{Cresci} G., {Mannucci} F., {Sommariva} V., {Maiolino} R., {Marconi} A.,
  {Brusa} M., 2012, \mnras, 421, 262

\bibitem[{{Cullen} {et~al}\mbox{.}(2014){Cullen}, {Cirasuolo}, {McLure},
  {Dunlop}, \& {Bowler}}]{cullen2014}
{Cullen} F., {Cirasuolo} M., {McLure} R.~J., {Dunlop} J.~S., {Bowler} R.~A.~A.,
  2014, \mnras, 440, 2300

\bibitem[{{Dalcanton}(2007)}]{dalcanton2007}
{Dalcanton} J.~J., 2007, \apj, 658, 941

\bibitem[{{Dalcanton}, {Yoachim} \& {Bernstein}(2004){Dalcanton}, {Yoachim}, \&
  {Bernstein}}]{dalcanton2004}
{Dalcanton} J.~J., {Yoachim} P., {Bernstein} R.~A., 2004, \apj, 608, 189

\bibitem[{{Dalla Vecchia} \& {Schaye}(2008)}]{dallavechia2008}
{Dalla Vecchia} C., {Schaye} J., 2008, \mnras, 387, 1431

\bibitem[{{Dav{\'e}}, {Finlator} \& {Oppenheimer}(2011){Dav{\'e}}, {Finlator},
  \& {Oppenheimer}}]{dave2010}
{Dav{\'e}} R., {Finlator} K., {Oppenheimer} B.~D., 2011, \mnras, 416, 1354

\bibitem[{{Dav{\'e}}, {Finlator} \& {Oppenheimer}(2012){Dav{\'e}}, {Finlator},
  \& {Oppenheimer}}]{dave2012}
{Dav{\'e}} R., {Finlator} K., {Oppenheimer} B.~D., 2012, \mnras, 421, 98

\bibitem[{{Davis} {et~al}\mbox{.}(1985){Davis}, {Efstathiou}, {Frenk}, \&
  {White}}]{davis1985}
{Davis} M., {Efstathiou} G., {Frenk} C.~S., {White} S.~D.~M., 1985, \apj, 292,
  371

\bibitem[{{Dayal}, {Ferrara} \& {Dunlop}(2013){Dayal}, {Ferrara}, \&
  {Dunlop}}]{dayal2012}
{Dayal} P., {Ferrara} A., {Dunlop} J.~S., 2013, \mnras, 430, 2891

\bibitem[{{De Rossi} {et~al}\mbox{.}(2013){De Rossi}, {Avila-Reese}, {Tissera},
  {Gonz{\'a}lez-Samaniego}, \& {Pedrosa}}]{derossi2013}
{De Rossi} M.~E., {Avila-Reese} V., {Tissera} P.~B., {Gonz{\'a}lez-Samaniego}
  A., {Pedrosa} S.~E., 2013, \mnras, 435, 2736

\bibitem[{{de Rossi}, {Tissera} \& {Scannapieco}(2007){de Rossi}, {Tissera}, \&
  {Scannapieco}}]{derossi2007}
{de Rossi} M.~E., {Tissera} P.~B., {Scannapieco} C., 2007, \mnras, 374, 323

\bibitem[{{Dolag} {et~al}\mbox{.}(2009){Dolag}, {Borgani}, {Murante}, \&
  {Springel}}]{dolag2009}
{Dolag} K., {Borgani} S., {Murante} G., {Springel} V., 2009, \mnras, 399, 497

\bibitem[{{Dressler} {et~al}\mbox{.}(1987){Dressler}, {Lynden-Bell},
  {Burstein}, {Davies}, {Faber}, {Terlevich}, \& {Wegner}}]{dressler1987}
{Dressler} A., {Lynden-Bell} D., {Burstein} D., {Davies} R.~L., {Faber} S.~M.,
  {Terlevich} R., {Wegner} G., 1987, \apj, 313, 42

\bibitem[{{Ellison} {et~al}\mbox{.}(2008){Ellison}, {Patton}, {Simard}, \&
  {McConnachie}}]{ellison2008}
{Ellison} S.~L., {Patton} D.~R., {Simard} L., {McConnachie} A.~W., 2008, \apjl,
  672, L107

\bibitem[{{Ellison} {et~al}\mbox{.}(2009){Ellison}, {Simard}, {Cowan},
  {Baldry}, {Patton}, \& {McConnachie}}]{ellison2009}
{Ellison} S.~L., {Simard} L., {Cowan} N.~B., {Baldry} I.~K., {Patton} D.~R.,
  {McConnachie} A.~W., 2009, \mnras, 396, 1257

\bibitem[{{Erb} {et~al}\mbox{.}(2006){Erb}, {Shapley}, {Pettini}, {Steidel},
  {Reddy}, \& {Adelberger}}]{erb2006}
{Erb} D.~K., {Shapley} A.~E., {Pettini} M., {Steidel} C.~C., {Reddy} N.~A.,
  {Adelberger} K.~L., 2006, \apj, 644, 813

\bibitem[{{Ferland} {et~al}\mbox{.}(1998){Ferland}, {Korista}, {Verner},
  {Ferguson}, {Kingdon}, \& {Verner}}]{ferland1998}
{Ferland} G.~J., {Korista} K.~T., {Verner} D.~A., {Ferguson} J.~W., {Kingdon}
  J.~B., {Verner} E.~M., 1998, \pasp, 110, 761

\bibitem[{{Finlator} \& {Dav{\'e}}(2008)}]{finlator2008}
{Finlator} K., {Dav{\'e}} R., 2008, \mnras, 385, 2181

\bibitem[{{Font} {et~al}\mbox{.}(2011){Font}, {McCarthy}, {Crain}, {Theuns},
  {Schaye}, {Wiersma}, \& {Dalla Vecchia}}]{font2011}
{Font} A.~S., {McCarthy} I.~G., {Crain} R.~A., {Theuns} T., {Schaye} J.,
  {Wiersma} R.~P.~C., {Dalla Vecchia} C., 2011, \mnras, 416, 2802

\bibitem[{{Gallazzi} {et~al}\mbox{.}(2005){Gallazzi}, {Charlot}, {Brinchmann},
  {White}, \& {Tremonti}}]{gallazzi2005}
{Gallazzi} A., {Charlot} S., {Brinchmann} J., {White} S.~D.~M., {Tremonti}
  C.~A., 2005, \mnras, 362, 41

\bibitem[{{Garnett} \& {Shields}(1987)}]{garnett1987}
{Garnett} D.~R., {Shields} G.~A., 1987, \apj, 317, 82

\bibitem[{{Gingold} \& {Monaghan}(1977)}]{gingold1977}
{Gingold} R.~A., {Monaghan} J.~J., 1977, \mnras, 181, 375

\bibitem[{{Haardt} \& {Madau}(2001)}]{haardt2001}
{Haardt} F., {Madau} P., 2001, in Clusters of Galaxies and the High Redshift
  Universe Observed in X-rays, {Neumann} D.~M., {Tran} J.~T.~V., eds.

\bibitem[{{Hayashi} {et~al}\mbox{.}(2009){Hayashi}, {Motohara}, {Shimasaku},
  {Onodera}, {Uchimoto}, {Kashikawa}, {Yoshida}, {Okamura}, {Ly}, \&
  {Malkan}}]{hayashi2009}
{Hayashi} M. {et~al.}, 2009, \apj, 691, 140

\bibitem[{{Henry} {et~al}\mbox{.}(2013){Henry}, {Martin}, {Finlator}, \&
  {Dressler}}]{henry2013}
{Henry} A., {Martin} C.~L., {Finlator} K., {Dressler} A., 2013, \apj, 769, 148

\bibitem[{{Hughes} {et~al}\mbox{.}(2013){Hughes}, {Cortese}, {Boselli},
  {Gavazzi}, \& {Davies}}]{hughes2013}
{Hughes} T.~M., {Cortese} L., {Boselli} A., {Gavazzi} G., {Davies} J.~I., 2013,
  \aap, 550, A115

\bibitem[{{Kennicutt}(1998)}]{kennicutt1998}
{Kennicutt}, Jr. R.~C., 1998, \apj, 498, 541

\bibitem[{{Kewley} \& {Ellison}(2008)}]{kewley2008}
{Kewley} L.~J., {Ellison} S.~L., 2008, \apj, 681, 1183

\bibitem[{{Kobayashi} \& {Nakasato}(2011)}]{Kobayashi11}
{Kobayashi} C., {Nakasato} N., 2011, \apj, 729, 16

\bibitem[{{Kobayashi}, {Springel} \& {White}(2007){Kobayashi}, {Springel}, \&
  {White}}]{kobayashi2007}
{Kobayashi} C., {Springel} V., {White} S.~D.~M., 2007, \mnras, 376, 1465

\bibitem[{{K{\"o}ppen} \& {Edmunds}(1999)}]{koppen1999}
{K{\"o}ppen} J., {Edmunds} M.~G., 1999, \mnras, 306, 317

\bibitem[{{K{\"o}ppen}, {Weidner} \& {Kroupa}(2007){K{\"o}ppen}, {Weidner}, \&
  {Kroupa}}]{koppen2007}
{K{\"o}ppen} J., {Weidner} C., {Kroupa} P., 2007, \mnras, 375, 673

\bibitem[{{Lamareille} {et~al}\mbox{.}(2009){Lamareille}, {Brinchmann},
  {Contini}, {Walcher}, {Charlot}, {P{\'e}rez-Montero}, {Zamorani}, {Pozzetti},
  {Bolzonella}, {Garilli}, {Paltani}, {Bongiorno}, {Le F{\`e}vre}, {Bottini},
  {Le Brun}, {Maccagni}, {Scaramella}, {Scodeggio}, {Tresse}, {Vettolani},
  {Zanichelli}, {Adami}, {Arnouts}, {Bardelli}, {Cappi}, {Ciliegi}, {Foucaud},
  {Franzetti}, {Gavignaud}, {Guzzo}, {Ilbert}, {Iovino}, {McCracken}, {Marano},
  {Marinoni}, {Mazure}, {Meneux}, {Merighi}, {Pell{\`o}}, {Pollo}, {Radovich},
  {Vergani}, {Zucca}, {Romano}, {Grado}, \& {Limatola}}]{lamareille2009}
{Lamareille} F. {et~al.}, 2009, \aap, 495, 53

\bibitem[{{Lamareille} {et~al}\mbox{.}(2004){Lamareille}, {Mouhcine},
  {Contini}, {Lewis}, \& {Maddox}}]{lamareille2004}
{Lamareille} F., {Mouhcine} M., {Contini} T., {Lewis} I., {Maddox} S., 2004,
  \mnras, 350, 396

\bibitem[{{Lara-L{\'o}pez} {et~al}\mbox{.}(2010){Lara-L{\'o}pez}, {Cepa},
  {Bongiovanni}, {P{\'e}rez Garc{\'{\i}}a}, {Ederoclite}, {Casta{\~n}eda},
  {Fern{\'a}ndez Lorenzo}, {Povi{\'c}}, \&
  {S{\'a}nchez-Portal}}]{laralopez2010a}
{Lara-L{\'o}pez} M.~A. {et~al.}, 2010, \aap, 521, L53

\bibitem[{{Lara-L{\'o}pez} {et~al}\mbox{.}(2013{\natexlab{a}}){Lara-L{\'o}pez},
  {Hopkins}, {L{\'o}pez-S{\'a}nchez}, {Brough}, {Colless}, {Bland-Hawthorn},
  {Driver}, {Foster}, {Liske}, {Loveday}, {Robotham}, {Sharp}, {Steele}, \&
  {Taylor}}]{laralopez2013c}
{Lara-L{\'o}pez} M.~A. {et~al.}, 2013{\natexlab{a}}, \mnras, 433, L35

\bibitem[{{Lara-L{\'o}pez} {et~al}\mbox{.}(2013{\natexlab{b}}){Lara-L{\'o}pez},
  {Hopkins}, {L{\'o}pez-S{\'a}nchez}, {Brough}, {Gunawardhana}, {Colless},
  {Robotham}, {Bauer}, {Bland-Hawthorn}, {Cluver}, {Driver}, {Foster},
  {Kelvin}, {Liske}, {Loveday}, {Owers}, {Ponman}, {Sharp}, {Steele}, {Taylor},
  \& {Thomas}}]{laralopez2013b}
{Lara-L{\'o}pez} M.~A. {et~al.}, 2013{\natexlab{b}}, \mnras, 434, 451

\bibitem[{{Lara-L{\'o}pez}, {L{\'o}pez-S{\'a}nchez} \&
  {Hopkins}(2013){Lara-L{\'o}pez}, {L{\'o}pez-S{\'a}nchez}, \&
  {Hopkins}}]{laralopez2013a}
{Lara-L{\'o}pez} M.~A., {L{\'o}pez-S{\'a}nchez} {\'A}.~R., {Hopkins} A.~M.,
  2013, \apj, 764, 178

\bibitem[{{Larson}(1974)}]{larson1974}
{Larson} R.~B., 1974, \mnras, 169, 229

\bibitem[{{Lee} {et~al}\mbox{.}(2006){Lee}, {Skillman}, {Cannon}, {Jackson},
  {Gehrz}, {Polomski}, \& {Woodward}}]{lee2006}
{Lee} H., {Skillman} E.~D., {Cannon} J.~M., {Jackson} D.~C., {Gehrz} R.~D.,
  {Polomski} E.~F., {Woodward} C.~E., 2006, \apj, 647, 970

\bibitem[{{Lequeux} {et~al}\mbox{.}(1979){Lequeux}, {Peimbert}, {Rayo},
  {Serrano}, \& {Torres-Peimbert}}]{lequeux1979}
{Lequeux} J., {Peimbert} M., {Rayo} J.~F., {Serrano} A., {Torres-Peimbert} S.,
  1979, \aap, 80, 155

\bibitem[{{Lilly} {et~al}\mbox{.}(2013){Lilly}, {Carollo}, {Pipino}, {Renzini},
  \& {Peng}}]{lilly2013}
{Lilly} S.~J., {Carollo} C.~M., {Pipino} A., {Renzini} A., {Peng} Y., 2013,
  \apj, 772, 119

\bibitem[{{Lucy}(1977)}]{lucy1977}
{Lucy} L.~B., 1977, \aj, 82, 1013

\bibitem[{{Maiolino} {et~al}\mbox{.}(2008){Maiolino}, {Nagao}, {Grazian},
  {Cocchia}, {Marconi}, {Mannucci}, {Cimatti}, {Pipino}, {Ballero}, {Calura},
  {Chiappini}, {Fontana}, {Granato}, {Matteucci}, {Pastorini}, {Pentericci},
  {Risaliti}, {Salvati}, \& {Silva}}]{maiolino2008}
{Maiolino} R. {et~al.}, 2008, \aap, 488, 463

\bibitem[{{Mannucci} {et~al}\mbox{.}(2010){Mannucci}, {Cresci}, {Maiolino},
  {Marconi}, \& {Gnerucci}}]{mannucci2010}
{Mannucci} F., {Cresci} G., {Maiolino} R., {Marconi} A., {Gnerucci} A., 2010,
  \mnras, 408, 2115

\bibitem[{{Mannucci} {et~al}\mbox{.}(2009){Mannucci}, {Cresci}, {Maiolino},
  {Marconi}, {Pastorini}, {Pozzetti}, {Gnerucci}, {Risaliti}, {Schneider},
  {Lehnert}, \& {Salvati}}]{mannucci2009}
{Mannucci} F. {et~al.}, 2009, \mnras, 398, 1915

\bibitem[{{Mannucci}, {Salvaterra} \& {Campisi}(2011){Mannucci}, {Salvaterra},
  \& {Campisi}}]{mannucci2011}
{Mannucci} F., {Salvaterra} R., {Campisi} M.~A., 2011, \mnras, 414, 1263

\bibitem[{{Marigo}(2001)}]{marigo2001}
{Marigo} P., 2001, \aap, 370, 194

\bibitem[{{McCarthy} {et~al}\mbox{.}(2012){McCarthy}, {Schaye}, {Font},
  {Theuns}, {Frenk}, {Crain}, \& {Dalla Vecchia}}]{mccarthy2012}
{McCarthy} I.~G., {Schaye} J., {Font} A.~S., {Theuns} T., {Frenk} C.~S.,
  {Crain} R.~A., {Dalla Vecchia} C., 2012, \mnras, 427, 379

\bibitem[{{Mouhcine} {et~al}\mbox{.}(2008){Mouhcine}, {Gibson}, {Renda}, \&
  {Kawata}}]{mouchine2008}
{Mouhcine} M., {Gibson} B.~K., {Renda} A., {Kawata} D., 2008, \aap, 486, 711

\bibitem[{{Moustakas} {et~al}\mbox{.}(2011){Moustakas}, {Zaritsky}, {Brown},
  {Cool}, {Dey}, {Eisenstein}, {Gonzalez}, {Jannuzi}, {Jones}, {Kochanek},
  {Murray}, \& {Wild}}]{moustakas2011}
{Moustakas} J. {et~al.}, 2011, arxiv: 1112.3300

\bibitem[{{Nomoto}, {Kobayashi} \& {Tominaga}(2013){Nomoto}, {Kobayashi}, \&
  {Tominaga}}]{Nomoto13}
{Nomoto} K., {Kobayashi} C., {Tominaga} N., 2013, \araa, 51, 457

\bibitem[{{Pasquali}, {Gallazzi} \& {van den Bosch}(2012){Pasquali},
  {Gallazzi}, \& {van den Bosch}}]{pasquali2012}
{Pasquali} A., {Gallazzi} A., {van den Bosch} F.~C., 2012, \mnras, 425, 273

\bibitem[{{Peeples} \& {Shankar}(2011)}]{peeples2011}
{Peeples} M.~S., {Shankar} F., 2011, \mnras, 417, 2962

\bibitem[{{Pilyugin} {et~al}\mbox{.}(2013){Pilyugin}, {Lara-L{\'o}pez},
  {Grebel}, {Kehrig}, {Zinchenko}, {L{\'o}pez-S{\'a}nchez}, {V{\'{\i}}lchez},
  \& {Mattsson}}]{pilyugin2013}
{Pilyugin} L.~S., {Lara-L{\'o}pez} M.~A., {Grebel} E.~K., {Kehrig} C.,
  {Zinchenko} I.~A., {L{\'o}pez-S{\'a}nchez} {\'A}.~R., {V{\'{\i}}lchez} J.~M.,
  {Mattsson} L., 2013, \mnras, 432, 1217

\bibitem[{{Pipino} {et~al}\mbox{.}(2009){Pipino}, {Chiappini}, {Graves}, \&
  {Matteucci}}]{pipino2009}
{Pipino} A., {Chiappini} C., {Graves} G., {Matteucci} F., 2009, \mnras, 396,
  1151

\bibitem[{{Portinari}, {Chiosi} \& {Bressan}(1998){Portinari}, {Chiosi}, \&
  {Bressan}}]{portinari1998}
{Portinari} L., {Chiosi} C., {Bressan} A., 1998, \aap, 334, 505

\bibitem[{{Romeo Velon{\`a}} {et~al}\mbox{.}(2013){Romeo Velon{\`a}},
  {Sommer-Larsen}, {Napolitano}, {Antonuccio-Delogu}, {Cielo}, {Gavignaud}, \&
  {Meza}}]{romeo2013}
{Romeo Velon{\`a}} A.~D., {Sommer-Larsen} J., {Napolitano} N.~R.,
  {Antonuccio-Delogu} V., {Cielo} S., {Gavignaud} I., {Meza} A., 2013, \apj,
  770, 155

\bibitem[{{Sales} {et~al}\mbox{.}(2012){Sales}, {Navarro}, {Theuns}, {Schaye},
  {White}, {Frenk}, {Crain}, \& {Dalla Vecchia}}]{sales2012}
{Sales} L.~V., {Navarro} J.~F., {Theuns} T., {Schaye} J., {White} S.~D.~M.,
  {Frenk} C.~S., {Crain} R.~A., {Dalla Vecchia} C., 2012, \mnras, 423, 1544

\bibitem[{{S{\'a}nchez} {et~al}\mbox{.}(2013){S{\'a}nchez}, {Rosales-Ortega},
  {Jungwiert}, {Iglesias-P{\'a}ramo}, {V{\'{\i}}lchez}, {Marino}, {Walcher},
  {Husemann}, {Mast}, {Monreal-Ibero}, {Cid Fernandes}, {P{\'e}rez},
  {Gonz{\'a}lez Delgado}, {Garc{\'{\i}}a-Benito}, {Galbany}, {van de Ven},
  {Jahnke}, {Flores}, {Bland-Hawthorn}, {L{\'o}pez-S{\'a}nchez}, {Stanishev},
  {Miralles-Caballero}, {D{\'{\i}}az}, {S{\'a}nchez-Blazquez}, {Moll{\'a}},
  {Gallazzi}, {Papaderos}, {Gomes}, {Gruel}, {P{\'e}rez}, {Ruiz-Lara},
  {Florido}, {de Lorenzo-C{\'a}ceres}, {Mendez-Abreu}, {Kehrig}, {Roth},
  {Ziegler}, {Alves}, {Wisotzki}, {Kupko}, {Quirrenbach}, {Bomans}, \& {Califa
  Collaboration}}]{sanchez2013}
{S{\'a}nchez} S.~F. {et~al.}, 2013, \aap, 554, A58

\bibitem[{{Santini} {et~al}\mbox{.}(2014){Santini}, {Maiolino}, {Magnelli},
  {Lutz}, {Lamastra}, {Li Causi}, {Eales}, {Andreani}, {Berta}, {Buat},
  {Cooray}, {Cresci}, {Daddi}, {Farrah}, {Fontana}, {Franceschini}, {Genzel},
  {Granato}, {Grazian}, {Le Floc'h}, {Magdis}, {Magliocchetti}, {Mannucci},
  {Menci}, {Nordon}, {Oliver}, {Popesso}, {Pozzi}, {Riguccini}, {Rodighiero},
  {Rosario}, {Salvato}, {Scott}, {Silva}, {Tacconi}, {Viero}, {Wang}, {Wuyts},
  \& {Xu}}]{santini2014}
{Santini} P. {et~al.}, 2014, \aap, 562, A30

\bibitem[{{Savaglio} {et~al}\mbox{.}(2005){Savaglio}, {Glazebrook}, {Le
  Borgne}, {Juneau}, {Abraham}, {Chen}, {Crampton}, {McCarthy}, {Carlberg},
  {Marzke}, {Roth}, {J{\o}rgensen}, \& {Murowinski}}]{savaglio2005}
{Savaglio} S. {et~al.}, 2005, \apj, 635, 260

\bibitem[{{Schaye}(2004)}]{schaye2004}
{Schaye} J., 2004, \apj, 609, 667

\bibitem[{{Schaye} {et~al}\mbox{.}(2015){Schaye}, {Crain}, {Bower}, {Furlong},
  {Schaller}, {Theuns}, {Dalla Vecchia}, {Frenk}, {McCarthy}, {Helly},
  {Jenkins}, {Rosas-Guevara}, {White}, {Baes}, {Booth}, {Camps}, {Navarro},
  {Qu}, {Rahmati}, {Sawala}, {Thomas}, \& {Trayford}}]{schaye2015}
{Schaye} J. {et~al.}, 2015, \mnras, 446, 521

\bibitem[{{Schaye} \& {Dalla Vecchia}(2008)}]{schaye2008}
{Schaye} J., {Dalla Vecchia} C., 2008, \mnras, 383, 1210

\bibitem[{{Schaye} {et~al}\mbox{.}(2010){Schaye}, {Dalla Vecchia}, {Booth},
  {Wiersma}, {Theuns}, {Haas}, {Bertone}, {Duffy}, {McCarthy}, \& {van de
  Voort}}]{schaye2010}
{Schaye} J. {et~al.}, 2010, \mnras, 402, 1536

\bibitem[{{Springel}(2005)}]{springel2005a}
{Springel} V., 2005, \mnras, 364, 1105

\bibitem[{{Springel} {et~al}\mbox{.}(2005){Springel}, {White}, {Jenkins},
  {Frenk}, {Yoshida}, {Gao}, {Navarro}, {Thacker}, {Croton}, {Helly},
  {Peacock}, {Cole}, {Thomas}, {Couchman}, {Evrard}, {Colberg}, \&
  {Pearce}}]{springel2005}
{Springel} V. {et~al.}, 2005, \nat, 435, 629

\bibitem[{{Springel}, {Yoshida} \& {White}(2001){Springel}, {Yoshida}, \&
  {White}}]{springel2001}
{Springel} V., {Yoshida} N., {White} S.~D.~M., 2001, \nat, 6, 79

\bibitem[{{Stewart} {et~al}\mbox{.}(2009){Stewart}, {Bullock}, {Wechsler}, \&
  {Maller}}]{stewart2009}
{Stewart} K.~R., {Bullock} J.~S., {Wechsler} R.~H., {Maller} A.~H., 2009, \apj,
  702, 307

\bibitem[{{Stott} {et~al}\mbox{.}(2013){Stott}, {Sobral}, {Bower}, {Smail},
  {Best}, {Matsuda}, {Hayashi}, {Geach}, \& {Kodama}}]{stott2013}
{Stott} J.~P. {et~al.}, 2013, \mnras, 436, 1130

\bibitem[{{Swinbank} {et~al}\mbox{.}(2012){Swinbank}, {Sobral}, {Smail},
  {Geach}, {Best}, {McCarthy}, {Crain}, \& {Theuns}}]{swinbank2012}
{Swinbank} A.~M., {Sobral} D., {Smail} I., {Geach} J.~E., {Best} P.~N.,
  {McCarthy} I.~G., {Crain} R.~A., {Theuns} T., 2012, \mnras, 426, 935

\bibitem[{{Theuns} {et~al}\mbox{.}(2002){Theuns}, {Schaye}, {Zaroubi}, {Kim},
  {Tzanavaris}, \& {Carswell}}]{theuns2002}
{Theuns} T., {Schaye} J., {Zaroubi} S., {Kim} T.-S., {Tzanavaris} P.,
  {Carswell} B., 2002, \apjl, 567, L103

\bibitem[{{Thielemann} {et~al}\mbox{.}(2003){Thielemann}, {Argast},
  {Brachwitz}, {Hix}, {H{\"o}flich}, {Liebend{\"o}rfer}, {Martinez-Pinedo},
  {Mezzacappa}, {Panov}, \& {Rauscher}}]{thielemann2003}
{Thielemann} F.-K. {et~al.}, 2003, Nuclear Physics A, 718, 139

\bibitem[{{Tissera}, {De Rossi} \& {Scannapieco}(2005){Tissera}, {De Rossi}, \&
  {Scannapieco}}]{tissera2005}
{Tissera} P.~B., {De Rossi} M.~E., {Scannapieco} C., 2005, \mnras, 364, L38

\bibitem[{{Tremonti} {et~al}\mbox{.}(2004){Tremonti}, {Heckman}, {Kauffmann},
  {Brinchmann}, {Charlot}, {White}, {Seibert}, {Peng}, {Schlegel}, {Uomoto},
  {Fukugita}, \& {Brinkmann}}]{tremonti2004}
{Tremonti} C.~A. {et~al.}, 2004, \apj, 613, 898

\bibitem[{{Troncoso} {et~al}\mbox{.}(2014){Troncoso}, {Maiolino}, {Sommariva},
  {Cresci}, {Mannucci}, {Marconi}, {Meneghetti}, {Grazian}, {Cimatti},
  {Fontana}, {Nagao}, \& {Pentericci}}]{troncoso2014}
{Troncoso} P. {et~al.}, 2014, \aap, 563, A58

\bibitem[{{Weinmann} {et~al}\mbox{.}(2012){Weinmann}, {Pasquali},
  {Oppenheimer}, {Finlator}, {Mendel}, {Crain}, \& {Macci{\`o}}}]{weinmann2012}
{Weinmann} S.~M., {Pasquali} A., {Oppenheimer} B.~D., {Finlator} K., {Mendel}
  J.~T., {Crain} R.~A., {Macci{\`o}} A.~V., 2012, \mnras, 426, 2797

\bibitem[{{Wiersma} {et~al}\mbox{.}(2010){Wiersma}, {Schaye}, {Dalla Vecchia},
  {Booth}, {Theuns}, \& {Aguirre}}]{wiersma2010}
{Wiersma} R.~P.~C., {Schaye} J., {Dalla Vecchia} C., {Booth} C.~M., {Theuns}
  T., {Aguirre} A., 2010, \mnras, 409, 132

\bibitem[{{Wiersma}, {Schaye} \& {Smith}(2009){Wiersma}, {Schaye}, \&
  {Smith}}]{wiersma2009a}
{Wiersma} R.~P.~C., {Schaye} J., {Smith} B.~D., 2009, \mnras, 393, 99

\bibitem[{{Wiersma}, {Schaye} \& {Theuns}(2011){Wiersma}, {Schaye}, \&
  {Theuns}}]{wiersma2011}
{Wiersma} R.~P.~C., {Schaye} J., {Theuns} T., 2011, \mnras, 415, 353

\bibitem[{{Wiersma} {et~al}\mbox{.}(2009){Wiersma}, {Schaye}, {Theuns}, {Dalla
  Vecchia}, \& {Tornatore}}]{wiersma2009b}
{Wiersma} R.~P.~C., {Schaye} J., {Theuns} T., {Dalla Vecchia} C., {Tornatore}
  L., 2009, \mnras, 399, 574

\bibitem[{{Yabe} {et~al}\mbox{.}(2014){Yabe}, {Ohta}, {Iwamuro}, {Akiyama},
  {Tamura}, {Yuma}, {Kimura}, {Takato}, {Moritani}, {Sumiyoshi}, {Maihara},
  {Silverman}, {Dalton}, {Lewis}, {Bonfield}, {Lee}, {Curtis-Lake}, {Macaulay},
  \& {Clarke}}]{yabe2013}
{Yabe} K. {et~al.}, 2014, \mnras, 437, 3647

\bibitem[{{Yabe} {et~al}\mbox{.}(2012){Yabe}, {Ohta}, {Iwamuro}, {Yuma},
  {Akiyama}, {Tamura}, {Kimura}, {Takato}, {Moritani}, {Sumiyoshi}, {Maihara},
  {Silverman}, {Dalton}, {Lewis}, {Bonfield}, {Lee}, {Curtis Lake}, {Macaulay},
  \& {Clarke}}]{yabe2012}
{Yabe} K. {et~al.}, 2012, \pasj, 64, 60

\bibitem[{{Yates}, {Kauffmann} \& {Guo}(2012){Yates}, {Kauffmann}, \&
  {Guo}}]{yates2011}
{Yates} R.~M., {Kauffmann} G., {Guo} Q., 2012, \mnras, 422, 215

\bibitem[{{Yuan}, {Kewley} \& {Richard}(2013){Yuan}, {Kewley}, \&
  {Richard}}]{yuan2013}
{Yuan} T.-T., {Kewley} L.~J., {Richard} J., 2013, \apj, 763, 9

\bibitem[{{Zahid} {et~al}\mbox{.}(2012){Zahid}, {Bresolin}, {Kewley}, {Coil},
  \& {Dav{\'e}}}]{zahid2012}
{Zahid} H.~J., {Bresolin} F., {Kewley} L.~J., {Coil} A.~L., {Dav{\'e}} R.,
  2012, \apj, 750, 120

\bibitem[{{Zahid} {et~al}\mbox{.}(2014){Zahid}, {Kashino}, {Silverman},
  {Kewley}, {Daddi}, {Renzini}, {Rodighiero}, {Nagao}, {Arimoto}, {Sanders},
  {Kartaltepe}, {Lilly}, {Maier}, {Geller}, {Capak}, {Carollo}, {Chu},
  {Hasinger}, {Ilbert}, {Kajisawa}, {Koekemoer}, {Kovac{\#728}}, {Le
  F{\`e}vre}, {Masters}, {McCracken}, {Onodera}, {Scoville}, {Strazzullo},
  {Sugiyama}, {Taniguchi}, \& {The COSMOS Team}}]{zahid2014}
{Zahid} H.~J. {et~al.}, 2014, \apj, 792, 75

\bibitem[{{Zaritsky}, {Kennicutt} \& {Huchra}(1994){Zaritsky}, {Kennicutt}, \&
  {Huchra}}]{zaritsky1994}
{Zaritsky} D., {Kennicutt}, Jr. R.~C., {Huchra} J.~P., 1994, \apj, 420, 87

\bibitem[{{Zhang} {et~al}\mbox{.}(2009){Zhang}, {Li}, {Kauffmann}, {Zou},
  {Catinella}, {Shen}, {Guo}, \& {Chang}}]{zhang2009}
{Zhang} W., {Li} C., {Kauffmann} G., {Zou} H., {Catinella} B., {Shen} S., {Guo}
  Q., {Chang} R., 2009, \mnras, 397, 1243

\end{thebibliography}

\section*{Appendix: Resolution study}

\begin{figure*}
\begin{center}
\resizebox{17cm}{!}{\includegraphics{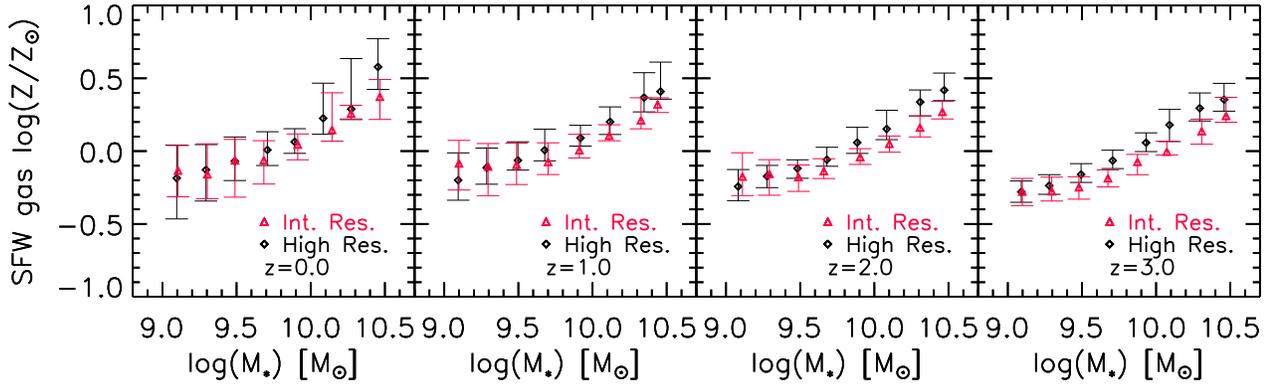}}
\end{center}
\caption[Zg vs Ms resolution]
{
SFR-weighted gas-phase metallicity as a function of stellar mass at $z \ge 0$ for
central galaxies in the high resolution run (black) and intermediate resolution run (red). 
The different curves with error bars depict the median relation with the 15th and 85th percentiles.
The level of numerical convergence is good.}
\label{fig:ap1}
\end{figure*}

\begin{figure*}
\begin{center}
\resizebox{17cm}{!}{\includegraphics{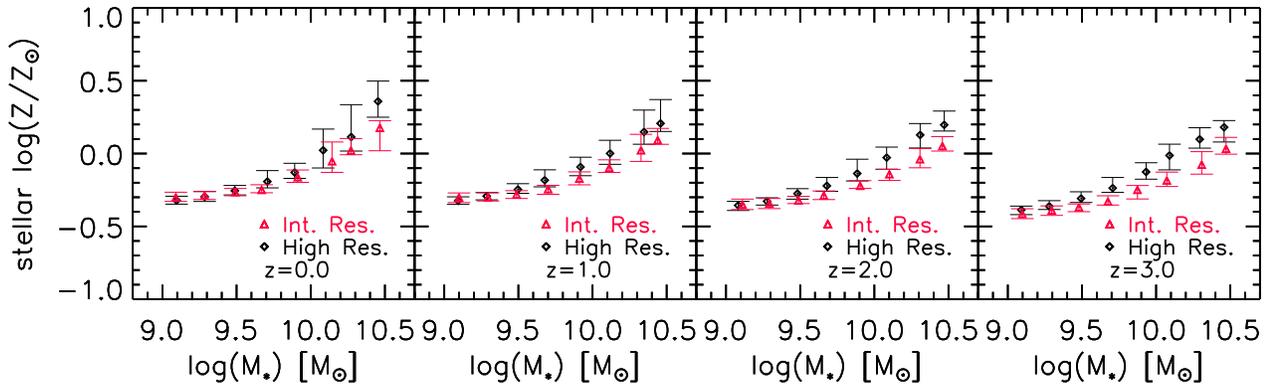}}
\end{center}
\caption[Zs vs Ms resolution]
{
Stellar metallicity as a function of stellar mass at $z \ge 0$ for
central galaxies in the high resolution run (black) and intermediate resolution run (red). 
The different curves with error bars depict the median relation with the 15th and 85th percentiles. Convergence is good, with a small tendency for higher values of $Z$ at higher resolution.}
\label{fig:ap2}
\end{figure*}

The analysis presented in this paper has been done by using the high resolution
version of \gimic\ simulations (see Section \ref{sec:simulation}).  In order to assess
the numerical convergence of our results, we have also performed a similar
analysis by using the intermediate resolution runs, which has 8 times coarser mass resolution.  
Our results are robust to a factor of 8 change in mass resolution

As an example, we compare the MZR based on SF-weighted gas and stellar metallicities at two resolutions, 
in Figs. \ref{fig:ap1} and \ref{fig:ap2}, respectively. We see that the main features of the relations seem to be robust against resolution. There is only a very small trend in the case of the intermediate resolution run to have lower abundances at the high-mass end towards higher redshifts.

\end{document}